\documentclass[aps,floats]{revtex4}
\usepackage{amsmath,amssymb}
\usepackage{graphicx,epsfig}
\usepackage[greek,english]{babel}
\usepackage{bbold}

\begin{document}
\bibliographystyle {plain}

\pdfoutput=1
\def\oppropto{\mathop{\propto}} 
\def\opsimeq{\mathop{\simeq}}
\def\opoverderline{\mathop{\overline}}
\def\operarrow{\mathop{\longrightarrow}}
\def\opsim{\mathop{\sim}}

\def\fig#1#2{\includegraphics[height=#1]{#2}}
\def\figx#1#2{\includegraphics[width=#1]{#2}}


\title{ Microcanonical conditioning of Markov processes on time-additive observables 
} 


\author{ C\'ecile Monthus }
 \affiliation{Universit\'e Paris Saclay, CNRS, CEA,
 Institut de Physique Th\'eorique, 
91191 Gif-sur-Yvette, France}

\begin{abstract}

The recent study by B. De Bruyne, S. N. Majumdar, H. Orland and G. Schehr [J. Stat. Mech. (2021) 123204], concerning the conditioning of the Brownian motion and of random walks on global dynamical constraints over a finite time-window $T$, is reformulated as a general framework for the 'microcanonical conditioning' of Markov processes on time-additive observables. This formalism is applied to various types of Markov processes, namely discrete-time Markov chains, continuous-time Markov jump processes and diffusion processes in arbitrary dimension. In each setting, the time-additive observable is also fully general, i.e. it can involve both the time spent in each configuration and the elementary increments of the Markov process. The various cases are illustrated via simple explicit examples. Finally, we describe the link with the 'canonical conditioning' based on the generating function of the time-additive observable for finite time $T$, while the regime of large time $T$ allows to recover the standard large deviation analysis of time-additive observables via the deformed Markov operator approach.

\end{abstract}

\maketitle


\section{ Introduction }

Time-additive observables of Markov processes have attracted a lot of interest recently,
in particular in the field of non equilibrium steady states in order to characterize their dynamical fluctuations over a large time-window $T$. From the point of view of the large deviation theory
(see the reviews \cite{oono,ellis,review_touchette} and references therein), 
time-additive observables belong to the Level 1 and can be thus analyzed via the contraction from higher Levels. 
For instance, the large deviations at Level 2 for the empirical density
allows to analyze the time-additive observables that only depend on the time spent in each configuration, but the Level 2 is usually not closed for non-equilibrium processes with steady currents. By contrast, 
the Level 2.5 concerning the joint distribution of the empirical density and of the empirical flows
 can be written in closed form for general Markov processes,
including discrete-time Markov chains
 \cite{fortelle_thesis,fortelle_chain,review_touchette,c_largedevdisorder,c_reset,c_inference},
continuous-time Markov jump processes
\cite{fortelle_thesis,fortelle_jump,maes_canonical,maes_onandbeyond,wynants_thesis,chetrite_formal,BFG1,BFG2,chetrite_HDR,c_ring,c_interactions,c_open,c_detailed,barato_periodic,chetrite_periodic,c_reset,c_inference,c_runandtumble,c_jumpdiff,c_skew,c_metastable,c_east,c_exclusion}
and Diffusion processes 
\cite{wynants_thesis,maes_diffusion,chetrite_formal,engel,chetrite_HDR,c_reset,c_lyapunov,c_inference,c_metastable}.
In addition, this Level 2.5 is necessary to analyze via contraction the general case of time-additive observables
that involve not only the time spent in each configuration but also the elementary increments of the Markov process. Another standard method to characterize the statistics of a time-additive observables is 
to study its generating function  
via the appropriate deformed Markov operator  that does not conserve the probability \cite{peliti,derrida-lecture,tailleur,sollich_review,lazarescu_companion,lazarescu_generic,jack_review,vivien_thesis,lecomte_chaotic,lecomte_thermo,lecomte_formalism,lecomte_glass,kristina1,kristina2,jack_ensemble,simon1,simon2,simon3,Gunter1,Gunter2,Gunter3,Gunter4,chetrite_canonical,chetrite_conditioned,chetrite_optimal,chetrite_HDR,touchette_circle,touchette_langevin,touchette_occ,touchette_occupation,derrida-conditioned,derrida-ring,bertin-conditioned,garrahan_lecture,Vivo,chemical,touchette-reflected,touchette-reflectedbis,c_lyapunov,previousquantum2.5doob,quantum2.5doob,quantum2.5dooblong,c_ruelle,lapolla,chabane}, while the probability-conserving Markov process corresponding to this 'canonical conditioning' can be written from the generalization of Doob's h-transform. 

On the other hand, the 'microcanonical conditioning' of one-dimensional stochastic processes 
on time-additive observables has been considered recently
in order to have efficient methods to generate stochastic trajectories satisfying global dynamical constraints over a finite time window $T$.
The conditioning on the area has been studied via various methods
for Brownian processes or bridges \cite{Mazzolo_Brownian} and for Ornstein-Uhlenbeck bridges \cite{Mazzolo_OU} 
(see also \cite{killing,Mazzolo_Taboo,Mazzolo_Strongly} for the discussion of other types of conditioning).
 The conditioning on the area and on other time-additive observables has been then analyzed 
 for the Brownian motion and for discrete-time random walks
\cite{bruyne_additive}, building on previous works  \cite{henri,satya,delarue,bruyne_discrete,bruyne_run,grela}
concerning the standard Doob conditioning, where the goal was to generate stochastic trajectories ending in a specific configuration at time $T$.

In the present paper, the approach of the recent study \cite{bruyne_additive} is reformulated 
as a general framework for the 'microcanonical conditioning' of Markov processes on time-additive observables, 
where the Markov process can be either a discrete-time Markov chain, a continuous-time Markov jump process
or a diffusion process in arbitrary dimension, while the time-additive observable
can involve both the time spent in each configuration and the increments of the Markov process.
This general formulation allows to make the link with the 'canonical conditioning' framework mentioned above.

The paper is organized as follows.
In section \ref{sec_gene}, we summarize the general ideas 
that allow to analyze the microcanonical conditioning of a Markov process
on a time-additive observable. 
The specific applications 
to discrete-time Markov chains, to continuous-time Markov jump processes
and to diffusion processes are then described in the sections \ref{sec_chain}, \ref{sec_jump} and \ref{sec_diff}
respectively.
Our conclusions are summarized in section \ref{sec_conclusion}.
The links with the canonical conditioning on a time-additive observable
are discussed for finite time $T$ in Appendix \ref{app_genecano} 
and for large $T$ in the large deviation regime for the time-additive observable in Appendix  \ref{app_largedev}.


\section{ Microcanonical conditioning on a time-additive observable  }

\label{sec_gene}

In this section, we summarize the general ideas and notations that will be useful in the whole paper.
 The equations will be written for discrete variables $(x,A)$, but 
the adaptation to continuous variables $(x,A)$ is of course straightforward : one just needs to replace sums by integrals, and discrete delta functions by continuous delta functions.

\subsection{ Notion of time-additive observable $A(t)$ for the Markov process $x(t)$ }

For the Markov process $x(t)$, the observable $A(t)$ is called time-additive if
the difference $\big( A(t_2)-A(t_1) \big)$ between the value $A(t_2)$ at time $t_2$ and the value $A(t_1)$ at time $t_1$
is a deterministic function ${\cal A}[.]$ of the Markov trajectory $x(t_1 \leq s \leq t_2)$ between $s=t_1$ and $s=t_2$
\begin{eqnarray}
A(t_2)-A(t_1)={\cal A}[x(t_1 \leq s \leq t_2) ]
\label{additivetraj}
\end{eqnarray}

\subsection{ Notion of microcanonical conditioning for the Markov process $x(t)$ and its time-additive observable $A(t)$  }

In the main text, we will focus on 
the 'microcanonical conditioning' where both the initial values $(x_0,A_0)$ at time $t=0$
and the final values $(x_T,A_T)$ at time $t=T$ are fixed for the Markov process $x(t)$ and its time-additive observable $A(t)$. In order to analyze what happens at intermediate times $t \in [0,T]$, 
the approach described in \cite {bruyne_additive} can be decomposed in the three steps described in the following three subsections.


\subsection{ Joint propagator $P_{t,t_0}(x,A \vert x_0,A_0)$ for the Markov process and its time-additive observable   }

\label{subsec_joint}

The first step concerns the joint propagator $P_{t,t_0}(x,A \vert x_0,A_0)$ 
of the Markov process $x(t)$ and of its time-additive observable $A(t)$
\begin{eqnarray}
P_{t,t_0}(x,A \vert x_0,A_0) \equiv \langle \delta_{x(t),x}\delta_{A(t),A}  \ \delta_{x(t_0),x_0} \delta_{A(t_0),A_0}\rangle
\label{jointpropa}
\end{eqnarray}
Since the time-additive observable is a deterministic function ${\cal A}[.]$ of the Markov trajectory $x(t_0 \leq s \leq t)$ (see Eq. \ref{additivetraj}), 
the joint propagator $P_{t,t_0}(x,A \vert x_0,A_0)$ satisfies :

(i) some Markov forward dynamics with respect to its final variables $(x,A) $ at time $t$

(ii) some Markov backward dynamics with respect to its initial variables $(x_0,A_0) $ at time $t_0$.


\subsection{ Conditional probability ${\cal P}^{Cond}_t(x,A) $ 
if starting at $(x_0,A_0)$ at time $t=0$ and ending at $(x_T,A_T)$ at time $t=T$ }

\label{subsec_cond}

The second step concerns the conditional probability ${\cal P}^{Cond}_t(x,A) $ 
to be at the values $(x,A)$ at some intermediate time $t \in ]0,T[$ 
if starting at the values $(x_0,A_0)$ at time $t=0$ and ending at the values $(x_T,A_T)$ at time $t=T$.
The probability $P_{T,0}(x_T,A_T \vert x_0,A_0)$ to end at $(x_T,A_T)$ at time $t=T$
when starting at $(x_0,A_0)$ at time $t=0$ satisfies
the Chapman-Kolmogorov equation with respect to any internal time $t \in ]0,T[$
\begin{eqnarray}
P_{T,0}(x_T,A_T \vert x_0,A_0) = \sum_x \sum_A P_{T,t}(x_T,A_T \vert x,A) P_{t,0}(x,A \vert x_0,A_0)
\label{chapman}
\end{eqnarray}
So
the conditional probability ${\cal P}^{Cond}_t(x,A)$ to see the values $(x,A)$ at the internal time $t \in ]0,T[$ 
is simply given by the ratio
\begin{eqnarray}
{\cal P}^{Cond}_t(x,A) =  \frac{P_{T,t}(x_T,A_T \vert x,A) P_{t,0}(x,A \vert x_0,A_0)}{P_{T,0}(x_T,A_T \vert x_0,A_0)}
\label{markovcond}
\end{eqnarray}
It is normalized as a consequence of Eq. \ref{chapman}
\begin{eqnarray}
\sum_x \sum_A {\cal P}^{Cond}_t(x,A) = 1
\label{markovcondnorma}
\end{eqnarray}
and it satisfies the fixed boundary conditions at time $t=0$ and at time $t=T$
\begin{eqnarray}
{\cal P}^{Cond}_{0}(x,A) && =  \frac{P_{T,0}(x_T,A_T \vert x,A) P_{0,0}(x,A \vert x_0,A_0)}{P_{T,0}(x_T,A_T \vert x_0,A_0)}
=   \delta_{x,x_0} \delta_{A,A_0}
\nonumber \\
{\cal P}^{Cond}_T(x,A) && =  \frac{P_{T,T}(x_T,A_T \vert x,A) P_{T,0}(x,A \vert x_0,A_0)}{P_{T,0}(x_T,A_T \vert x_0,A_0)}
 = \delta_{x,x_T}  \delta_{A,A_T}
\label{markovcondboundary}
\end{eqnarray}


\subsection{ Markov dynamics for the conditional probability ${\cal P}^{Cond}_t(x,A) $  }

The third step consists in deriving the 
Markov dynamics of the conditional probability ${\cal P}^{Cond}_t(x,A) $
 from the Markov dynamics satisfied by the two joints propagators in the numerator of Eq. \ref{markovcond},
 namely :

(i) the Markov forward dynamics of the joint propagator $P_{t,0}(x,A \vert x_0,A_0)$ with respect to its final variables $(x,A) $ at time $t$

(ii) the Markov backward dynamics of the joint propagator $P_{T,t}(x_T,A_T \vert x,A) $ with respect to its initial variables $(x,A) $ at time $t$

In the three following sections,
the Markov dynamics for the conditional probability ${\cal P}^{Cond}_t(x,A) $ 
is written explicitly for discrete-time Markov chains (section \ref{sec_chain}), 
for continuous-time Markov jump processes (section \ref{sec_jump})
and for diffusion processes (\ref{sec_diff}).


\section{ Application to discrete-time Markov chains  }

\label{sec_chain}

In this section, we focus 
on the Markov Chain dynamics where the probability $P_t(x)  $ to be in the configuration $x$ at time $t$
evolves according to
\begin{eqnarray}
P_{t+1}(x) =  \sum_{x'} W(x ; x')  P_t(x')
\label{markovchain}
\end{eqnarray}
The matrix element $W(x ; x')  \in [0,1]$ represents the probability 
to be in the configuration $x$ at time $(t+1)$ if in the configuration $x'$ at $t$,
with the normalization for any $x'$
\begin{eqnarray}
  \sum_x W(x ; x')  =1
\label{markovnorma0}
\end{eqnarray}

The time-additive observable $A(t)$ of the trajectory $x(t_1 \leq s \leq t_2)$ of Eq. \ref{additivetraj}
can be parametrized by some function $\beta(x,y)$
\begin{eqnarray}
 A(t_2)-A(t_1)={\cal A}[x(t_1 \leq s \leq t_2) ]   \equiv       \sum_{s=t_1+1}^{t_2} \beta(x(s),x(s-1))
\label{additivechain}
\end{eqnarray}

Since the time $t$ and the space $x$ are both discrete, the equations will be written below for 
the case of a discrete variable $A$, but 
the adaptation to a continuous variable $A$ is of course straightforward :
one just needs to replace sums by integrals, and discrete delta functions by continuous delta functions.


\subsection{ Dynamics of the joint propagator $P_{t,t_0}(x,A \vert x_0,A_0)$   }

Since the increment between $t$ and $(t+1)$ of the time-additive observable $A(t)$ of Eq. \ref{additivechain}
 reduces to
\begin{eqnarray}
 A(t+1)-A(t)  =    \beta(x(t+1),x(t))
\label{additivechainelementary}
\end{eqnarray}
one just needs to introduce the joint generator
\begin{eqnarray}
W(x,A ; x', A') = W(x ; x') \delta_{A, A'  + \beta(x,x')} 
\label{markovmatrixA}
\end{eqnarray}
that involves the initial Markov matrix $W(x ; x') $ of Eq. \ref{markovchain},
while the delta function in $A$ describes the deterministic evolution of the time-additive observable
once the configurations $x$ and $x'$ are given.
The normalization of Eq. \ref{markovnorma0}
ensures the normalization of the joint generator 
 for any $(x',A')$
\begin{eqnarray}
  \sum_x \sum_A W(x,A ; x', A')   =\sum_x W(x ; x')  =1 
\label{markovnorma}
\end{eqnarray}
The joint propagator $P_{t,t_0}(x,A \vert x_0,A_0)$ of Eq. \ref{jointpropa} satisfies

(i) the forward dynamics with respect to the final variables $(x,A) $
\begin{eqnarray}
P_{t+1,t_0}(x,A\vert x_0 A_0) =  \sum_{x'} \sum_{A'} W(x,A ; x', A')  P_{t,t_0}(x',A'\vert x_0 A_0)
\label{markovchainAforward}
\end{eqnarray}

(ii) the backward dynamics with respect to the initial variables $(x_0,A_0) $ 
\begin{eqnarray}
P_{t,t_0-1}(x,A\vert x_0 A_0) =  \sum_{x_0'} \sum_{A_0'}  P_{t,t_0}(x,A\vert x_0' A_0')W(x_0',A_0' ; x_0, A_0)
\label{markovchainAbackward}
\end{eqnarray}


\subsection{ Forward Markov dynamics for the conditional probability ${\cal P}^{Cond}_t(x,A) $ with a time-dependent generator }

Let us plug the forward dynamics of Eq. \ref{markovchainAforward} for $P_{t+1}(x,A \vert x_0,A_0) $
into the conditional probability of Eq. \ref{markovcond} at time $(t+1)$
\begin{eqnarray}
{\cal P}^{Cond}_{t+1}(x,A) && =  \frac{P_{T,t+1}(x_T,A_T \vert x,A) }{P_{T,0}(x_T,A_T \vert x_0,A_0)} P_{t+1,0}(x,A \vert x_0,A_0)
\nonumber \\
&& 
 =  \frac{P_{T,t+1}(x_T,A_T \vert x,A) }{P_{T,0}(x_T,A_T \vert x_0,A_0)}
 \sum_{x'} \sum_{A'} W(x,A ; x', A')  P_{t,0}(x',A'\vert x_0 A_0)
\label{markovcondnextiter}
\end{eqnarray}
Let us now use the conditional probability at time $t$ of Eq. \ref{markovcond} to replace $P_{t,t_0}(x',A'\vert x_0 A_0) $ 
\begin{eqnarray}
 P_{t,0}(x',A' \vert x_0,A_0)
 = \frac{P_{T,0}(x_T,A_T \vert x_0,A_0)}{P_{T,t}(x_T,A_T \vert x',A')} {\cal P}^{Cond}_t(x',A')  
\label{markovcondprime}
\end{eqnarray}
in order
to rewrite Eq. \ref{markovcondnextiter} as the forward Markov dynamics
\begin{eqnarray}
{\cal P}^{Cond}_{t+1}(x,A) && = 
 \frac{P_{T,t+1}(x_T,A_T \vert x,A) }{P_{T,0}(x_T,A_T \vert x_0,A_0)}
 \sum_{x'} \int d A' W(x,A ; x', A')  \frac{P_{T,0}(x_T,A_T \vert x_0,A_0)}{P_{T,t}(x_T,A_T \vert x',A')} {\cal P}^{Cond}_t(x',A')  
 \nonumber \\
 && \equiv  \sum_{x'} \int d A' W^{Forw[x_T,A_T;T]}_{t+1/2}(x,A ; x', A') {\cal P}^{Cond}_t(x',A') 
\label{markovconddyn}
\end{eqnarray}
where the generator  
associated to this forward conditioned dynamics 
\begin{eqnarray}
W^{Forw[x_T,A_T;T]}_{t+1/2}(x,A ; x', A') \equiv P_{T,t+1}(x_T,A_T \vert x,A) W(x,A ; x', A')  \frac{ 1}{P_{T,t}(x_T,A_T \vert x',A')} 
\label{markovcondforward}
\end{eqnarray}
is time-dependent because the joint generator $W(x,A ; x', A') $ of Eq. \ref{markovmatrixA}
is conjugated with the full propagators 
$P_{T,t+1}(x_T,A_T \vert x,A) $ and $P_{T,t}(x_T,A_T \vert x',A') $ up to the imposed final values $(x_T,A_T)$ at time $T$.
The normalization for any $(x',A')$ of this conditional forward generator
\begin{eqnarray}
\sum_x \sum_A W^{Forw[x_T,A_T;T]}_{t+1/2}(x,A ; x', A') = 1 
\label{markovconddynnorma}
\end{eqnarray}
is ensured by the backward recursion of Eq. \ref{markovchainAbackward}.

The physical meaning of the generator of Eq. \ref{markovcondforward}
is that in the conditioned dynamics,
  the possibles transitions are the same as in the initial dynamics
(an impossible transition $W(x,A ; x', A')=0 $ in the initial dynamics remains impossible 
$W^{Forw[x_T,A_T;T]}_{t+1/2}(x,A ; x', A')=0$ in the conditioned dynamics), but the possible transitions
have different probabilities that have changed from $W(x,A ; x', A') $ to $W^{Forw[x_T,A_T;T]}_{t+1/2}(x,A ; x', A') $.

In practice, if one wishes to use these new probabilities $W^{Forw[x_T,A_T;T]}_{t+1/2}(x,A ; x', A') $
to generate stochastic trajectories of the conditioned dynamics, one needs to know
the explicit form of the joint propagator $P_{t,t_0}(x,A \vert x_0,A_0)$ of Eq. \ref{jointpropa} 
satisfying the joint forward dynamics of Eq. \ref{markovchainAforward}.


\subsection{ Backward Markov dynamics for the conditional probability ${\cal P}^{Cond}_t(x,A) $ with a time-dependent generator }

Let us write the backward recursion of Eq. \ref{markovchainAbackward}
for $P_{T,t}(x_T,A_T \vert x,A) $
\begin{eqnarray}
P_{T,t}(x_T,A_T \vert x,A) = \sum_{x'} \sum_{A'} P_{T,t+1}(x_T,A_T \vert x',A')W(x',A' ; x, A)
\label{markovchainAbackwardfuture}
\end{eqnarray}
and use the conditional probability of Eq. \ref{markovcond} at time $(t+1)$ to make the replacement
\begin{eqnarray}
P_{T,t+1}(x_T,A_T \vert x',A') 
  = {\cal P}^{Cond}_{t+1}(x',A') \frac{P_{T,0}(x_T,A_T \vert x_0,A_0)}{P_{t+1,0}(x',A' \vert x_0,A_0)}
\label{markovcondnextbis}
\end{eqnarray}
into order to rewrite the conditional probability of Eq. \ref{markovcond} as
\begin{eqnarray}
{\cal P}^{Cond}_t(x,A) && =  \sum_{x'} \sum_{A'} P_{T,t+1}(x_T,A_T \vert x',A')
\frac{ P_{t,0}(x,A \vert x_0,A_0)}{P_{T,0}(x_T,A_T \vert x_0,A_0)}
\nonumber \\
&& =  \sum_{x'} \sum_{A'} {\cal P}^{Cond}_{t+1}(x',A') \frac{P_{T,0}(x_T,A_T \vert x_0,A_0)}{P_{t+1,0}(x',A' \vert x_0,A_0)}
W(x',A' ; x, A)  \frac{ P_{t,0}(x,A \vert x_0,A_0)}{P_{T,0}(x_T,A_T \vert x_0,A_0)}
\nonumber \\
&& \equiv   \sum_{x'} \sum_{A'} {\cal P}^{Cond}_{t+1}(x',A') W^{Backw[x_0,A_0;0]}_{t+1/2} (x',A' ; x, A)
\label{markovconddynback}
\end{eqnarray}
where the generator  
associated to this backward conditioned dynamics 
\begin{eqnarray}
W^{Backw[x_0,A_0;0]}_{t+1/2} (x',A' ; x, A) \equiv \frac{ 1 }{P_{t+1,0}(x',A' \vert x_0,A_0)}
W(x',A' ; x, A)   P_{t,0}(x,A \vert x_0,A_0)
\label{markovcondbackward}
\end{eqnarray}
involves the conjugation of joint generator $W(x',A' ; x, A)   $ of Eq. \ref{markovmatrixA}
 by the full propagators 
$P_{t+1,0}(x',A' \vert x_0,A_0)$ and $  P_{t,0}(x,A \vert x_0,A_0) $ up to the imposed initial values $(x_0,A_0)$ at time $t=0$.


\subsection{ Illustration with a simple example : conditioning the Sisyphus Random Walk on the number of resets  }

\label{subsec_resetchain}

In the field of stochastic resetting (see the review \cite{review_reset} and references therein),
one of the simplest example is the Sisyphus Random Walk \cite{sisyphus} defined on the semi-infinite lattice $x=0,1,2,..$ with the Markov matrix
\begin{eqnarray}
W(x ; x')  = R \delta_{x,0}  +  (1-R) \delta_{x,x'+1}        
\label{wxy1d}
\end{eqnarray}
The physical meaning is that
when Sisyphus is at position $x$ at time $t$, he 
can either return to the origin $x=0$ with the reset probability $R \in ]0,1[$ or
he can move forward to the next position $(x+1)$ with the complementary probability $(1-R)$.

As time-additive observable of Eq. \ref{additivechain}, we will choose 
the function
\begin{eqnarray}
\beta(x,x') = \delta_{x,0}
\label{betachoice}
\end{eqnarray}
in order to count the total number of resets to the origin during the time interval 
\begin{eqnarray}
A(t_2)-A(t_1)={\cal A}[x(t_1 \leq s \leq t_2) ]   \equiv       \sum_{s=t_1+1}^{t_2} \delta_{x(s),0}
\label{additivechainreset}
\end{eqnarray}

\subsubsection{ Explicit form of the joint propagator $P_{t,t_0}(x,A \vert x_0 A_0) $  }

The joint generator of Eq. \ref{markovmatrixA} 
\begin{eqnarray}
W(x ,A ; x', A')  =  W(x ; x') \delta_{A, A'  + \delta_{x,0}} = 
R \delta_{x,0} \delta_{A,A'+1}      +  (1-R) \delta_{x,x'+1}  \delta_{A,A'}      
\label{wxy1da}
\end{eqnarray}
governs the forward dynamics of Eq. \ref{markovchainAforward} for the joint propagator
\begin{eqnarray}
P_{t+1,t_0}(x,A \vert x_0 A_0) && =  \sum_{x'=0}^{+\infty}  \sum_{A'=A_0}^{A}      W(x ,A ; x', A')    
P_{t,t_0}(x',A' \vert x_0 A_0)
\nonumber \\
&& =  
  R \delta_{x,0} \theta( A>A_0) \sum_{x'=0}^{+\infty}  P_{t,t_0}(x',A-1 \vert x_0 A_0)
+ \theta(x >0)        (1-R)      
P_{t,t_0}(x-1,A \vert x_0 A_0)
\label{markovchainreset1djoint}
\end{eqnarray}
where the notation $\theta $  is used to denote the inequalities that need to be satisfied.
The solution can be directly written from the renewal analysis of the dynamics 
\begin{eqnarray}
&& P_{t,t_0}(x,A \vert x_0 A_0)  = (1-R)^{(t-t_0)} \delta_{A,A_0} \delta_{x,x_0+(t-t_0)}
\nonumber \\
&& + R^{A-A_0} (1-R)^{(t-t_0)-(A-A_0)} \frac{ \big[ (t-t_0) -x-1 \big]! }
{\big[ (A-A_0) -1 \big]!  \big[(t-t_0) -(A-A_0)-x \big]!} \theta(1 \leq A-A_0 \leq (t-t_0) -x )
\label{jointpropagatorsolution}
\end{eqnarray}
The summation of Eq. \ref{jointpropagatorsolution}
over the variable $x$ allows to recover that the propagator for the variable $A$ alone corresponds to the binomial distribution for $(A-A_0)$ and is independent of $x_0$
\begin{eqnarray}
 \sum_{x=0}^{+\infty} P_{t,t_0}(x,A \vert x_0 A_0) && = 
 R^{A-A_0} (1-R)^{(t-t_0)-(A-A_0)} \frac{  (t-t_0)  ! }
{ (A-A_0)  !  \big[(t-t_0) -(A-A_0) \big]!} \theta(0 \leq A-A_0 \leq (t-t_0)  )
\nonumber \\
&& \equiv P_{t,t_0}(A \vert A_0)
\label{jointpropagatorsolutionAalone}
\end{eqnarray}
The summation of Eq. \ref{jointpropagatorsolution}
over the variable $A$ yields the propagator for the initial Sisyphus random walk $x(t)$ alone
\begin{eqnarray}
 \sum_{A=A_0}^{+\infty} P_{t,t_0}(x,A \vert x_0 A_0) && = (1-R)^{(t-t_0)}  \delta_{x,x_0+(t-t_0)}
+ R (1-R)^x  \theta( 0 \leq x \leq (t-t_0) -1 )
  \equiv P_{t,t_0}(x \vert x_0)
\label{jointpropagatorsolutionxalone}
\end{eqnarray}
that converges towards the steady state corresponding to the geometric distribution
\begin{eqnarray}
P_{t,t_0}(x \vert x_0) \opsimeq_{(t-t_0) \to +\infty}  R (1-R)^x  \theta( 0 \leq x  )  \equiv P_{st}(x )
\label{jointpropagatorsolutionxalonesteady}
\end{eqnarray}


\subsubsection{ Forward generator of the conditioned dynamics : resetting probabilities depending on time and configuration   }

For the conditioned dynamics, the 
 forward generator of Eq. \ref{markovcondforward}
reads using the joint generator of Eq. \ref{wxy1d}
\begin{eqnarray}
 W^{Forw[x_T,A_T;T]}_{t+1/2}(x,A ; x', A') && = P_{T,t+1}(x_T,A_T \vert x,A) W(x,A ; x', A')  \frac{ 1}{P_{T,t}(x_T,A_T \vert x',A')} 
\nonumber \\
&& 
= R \delta_{x,0} \delta_{A,A'+1}  \frac{ P_{T,t+1}(x_T,A_T \vert 0,A)}{P_{T,t}(x_T,A_T \vert x',A')} 
    + (1-R) \delta_{x,x'+1}  \delta_{A,A'}    \frac{ P_{T,t+1}(x_T,A_T \vert x,A)}{P_{T,t}(x_T,A_T \vert x',A)}   
 \nonumber \\
&& 
\equiv  \delta_{x,0} \delta_{A,A'+1}  R^{Forw[x_T,A_T;T]}_{t+1/2}(x',A') 
    +  \delta_{x,x'+1}  \delta_{A,A'}    \left( 1- R^{Forw[x_T,A_T;T]}_{t+1/2}(x',A')  \right)      
    \ \ 
\label{markovcondforwardreset}
\end{eqnarray}
where the effective resetting probability towards the origin
$x=0$ depends on the time $t$ and on the values $(x',A')$ at time $t$ via
\begin{eqnarray}
R^{Forw[x_T,A_T;T]}_{t+1/2}(x',A') = R   \frac{ P_{T,t+1}(x_T,A_T \vert 0,A'+1)}{P_{T,t}(x_T,A_T \vert x',A')}   
\label{effectiveresetproba}
\end{eqnarray}
One can plug the explicit form of Eq. \ref{jointpropagatorsolution}
for the joint propagator to obtain the explicit form of
the effective resetting probability of Eq. \ref{markovcondforwardreset}.

In summary, the conditioned dynamics corresponds to a Sisyphus random walk with modified resetting probabilities:
when Sisyphus is in the configuration $(x',A')$ at time $t$, he 
can either return to the origin $x=0$ and increment the observable $A=A'+1$ 
with the reset probability $R^{Forw[x_T,A_T;T]}_{t+1/2}(x',A') $ or
he can move forward to the next position $(x'+1)$ and keep the observable $A=A'$ with the complementary probability $[1-R^{Forw[x_T,A_T;T]}_{t+1/2}(x',A')]$.

The 'canonical conditioning' (see the reminder in the two Appendices) of the Sisyphus Random Walk 
has been studied in \cite{c_reset} for the more general case where the reset probabilities of the initial model are space-dependent $R_x$ (instead of being given by the constant value $R$)
and where the time-additive observable involves an arbitrary function $\beta(x,x')$.

As a final remark, let us stress that other explicit examples of microcanonical conditioning for discrete-time random walks on time-additive observables
can be found in \cite{bruyne_additive}.


\section{ Application to continuous-time Markov jump processes   }

\label{sec_jump}

In this section, we consider the continuous-time dynamics in discrete configuration space defined by the Master Equation
\begin{eqnarray}
\partial_t P_t(x)  =    \sum_{x' \ne x}  \left[ w(x;x') P_t(x')  - w(x';x) P_t(x) \right]
\label{mastereq}
\end{eqnarray}
where $w(x;x') \geq 0$ represents the transition rate from $x'$ towards $x \ne x' $.

The time-additive observable $A(t)$ of the trajectory $x(t_1 \leq s \leq t_2)$ of Eq. \ref{additivetraj}
can be parametrized by the two functions $\alpha(x)$ and $\beta(x,y)$
\begin{eqnarray}
A(t_2)-A(t_1)={\cal A}[x(t_1 \leq s \leq t_2) ]=  \int_{t_1}^{t_2} ds \alpha(x(s))  +   \sum_{s \in [t_1,t_2]: x(s^+) \ne x(s) } \beta(x(s^+),x(s))
\label{additivejump}
\end{eqnarray}
Whenever the function $\alpha(.)$ is present, the observable $A$ is continuous, so
the equations will be written for continuous $A$ in the following general subsections,
while an example with discrete variable $A$ will be given in the last subsection.


\subsection{  Dynamics of the joint propagator $P_{t,t_0}(x,A \vert x_0,A_0)$  }

Between $t$ and $(t+dt)$, the elementary increment of the time-additive observable of Eq. \ref{additivejump}
reduces to
\begin{eqnarray}
 A(t+dt)-A(t)   =    dt \alpha(x(t))  +   \delta_{x(t+dt) \ne x(t) } \beta(x(t+dt),x(t))
\label{additivejumpincrement}
\end{eqnarray}
As a consequence, the function $\alpha(.)$ corresponds to a deterministic drift for the continuous 
observable $A$,
while the function $\beta(.,.) $ will appear via the following delta function 
in the joint jump rates from $x'$ to $x \ne x'$
\begin{eqnarray}
w(x,A;x',A')  \equiv  w(x;x') \delta(A-A'- \beta(x,x') )
\label{wjoint}
\end{eqnarray}
So the joint propagator $P_{t,t_0}(x,A \vert x_0,A_0)$ of Eq. \ref{jointpropa} satisfies :

(i) the forward jump-drift dynamics with respect to the final values $(x,A)$ at time $t$
\begin{eqnarray}
\partial _t P_{t,t_0}(x,A \vert x_0,A_0) 
&& = - \partial_A \left[ \alpha(x) P_{t,t_0}(x,A\vert x_0,A_0) \right] 
\nonumber \\ &&   
+  \sum_{x' \ne x } \int d A' \bigg[  w(x,A;x',A')  P_{t,t_0}(x',A' \vert x_0,A_0) -  w(x',A';x,A)P_{t,t_0}(x,A\vert x_0,A_0)\bigg]
\label{markovjumpAforward}
\end{eqnarray}

(ii) the backward jump-drift dynamics with respect to the initial values $(x_0,A_0)$ at time $t_0$
\begin{eqnarray}
- \partial _{t_0} P_{t,t_0}(x,A \vert x_0,A_0) 
&& = \alpha(x_0) \partial_{A_0}  P_{t,t_0}(x,A\vert x_0,A_0) 
\nonumber \\
&& + 
 \sum_{x_0' \ne x_0 } \int d A_0'  
\bigg [P_{t,t_0}(x,A\vert x_0',A_0')  -P_{t,t_0}(x,A\vert x_0,A_0) \bigg]
w(x_0',A_0';x_0,A_0)  
\label{markovjumpAbackward}
\end{eqnarray}


\subsection{ Markov dynamics for the conditional probability ${\cal P}^{Cond}_t(x,A) $ with a time-dependent generator }

Let us now focus on the dynamics for the conditional probability ${\cal P}^{Cond}_t(x,A) $ of Eq. \ref{markovcond}.
Its time-derivative involves the derivatives of the two propagators of the numerator
\begin{eqnarray}
\partial_t  {\cal P}^{Cond}_t(x,A) && = \left[ \partial_t P_{T,t}(x_T,A_T \vert x,A)\right] 
\frac{ P_{t,0}(x,A \vert x_0,A_0)}{P_{T,0}(x_T,A_T \vert x_0,A_0)}
+  \frac{P_{T,t}(x_T,A_T \vert x,A) }{P_{T,0}(x_T,A_T \vert x_0,A_0)} 
\left[ \partial_t P_{t,0}(x,A \vert x_0,A_0) \right]
\label{markovcondjumpderi}
\end{eqnarray}
Since the propagator $P_{t,0}(x,A \vert x_0,A_0) $ 
satisfies the forward dynamics of Eq. \ref{markovjumpAforward},
and
since the propagator $P_{T,t}(x_T,A_T \vert x,A) $
satisfies the backward dynamics of Eq. \ref{markovjumpAbackward}
\begin{eqnarray}
-\partial _t P_{T,t}(x_T,A_T \vert x,A) 
&& = \alpha(x) \partial_{A}  P_{T,t}(x_T,A_T\vert x,A) 
\nonumber \\
&& + 
 \sum_{x' \ne x } \int d A'  \left[ P_{T,t}(x_T,A_T \vert  x',A') - P_{T,t}(x_T,A_T\vert x,A) \right]  w(x',A';x,A)  
\label{markovjumpAbackwardbis}
\end{eqnarray}
 Eq. \ref{markovcondjumpderi}
becomes
\begin{eqnarray}
 \partial_t  {\cal P}^{Cond}_t(x,A) 
&& =     
  - \alpha(x)  \frac{ P_{t,0}(x,A \vert x_0,A_0)}{P_{T,0}(x_T,A_T \vert x_0,A_0)}  \partial_{A}  P_{T,t}(x_T,A_T\vert x,A) 
 - \alpha(x) \frac{P_{T,t}(x_T,A_T \vert x,A) }{P_{T,0}(x_T,A_T \vert x_0,A_0)}  \partial_A   P_{t,0}(x,A\vert x_0,A_0)
\nonumber \\ &&
- \frac{ P_{t,0}(x,A \vert x_0,A_0)}{P_{T,0}(x_T,A_T \vert x_0,A_0)} \sum_{x' \ne x } \int d A'  P_{T,t}(x_T,A_T \vert  x',A')   w(x',A';x,A) 
\nonumber \\ &&
+  \frac{P_{T,t}(x_T,A_T \vert x,A) }{P_{T,0}(x_T,A_T \vert x_0,A_0)} 
  \sum_{x' \ne x } \int d A'   w(x,A;x',A')  P_{t,0}(x',A' \vert x_0,A_0) 
 \nonumber \\ &&  
\label{markovcondjumpdericalcul}
\end{eqnarray}

(i) Forward perspective : Eq. \ref{markovcond} allows to replace all the propagators on $[0,t]$
\begin{eqnarray}
P_{t,0}( x,A \vert  x_0,A_0) =
 {\cal P}^{Cond}_t( x,A) \frac{P_{T,0}( x_T,A_T \vert  x_0,A_0)}{P_{T,t}( x_T,A_T \vert  x,A)}
\label{markovcondelimpt}
\end{eqnarray}
in Eq. \ref{markovcondjumpdericalcul} to obtain the forward dynamics
\begin{eqnarray}
&& \partial_t  {\cal P}^{Cond}_t(x,A) 
 =    - \alpha(x)   \partial_A   {\cal P}^{Cond}_t( x,A)
\nonumber \\ &&
+  
  \sum_{x' \ne x } \int d A'  \bigg[ w_t^{Forw[x_T,A_T;T]} (x,A ; x', A')   {\cal P}^{Cond}_t( x',A') 
  - w_t^{Forw[x_T,A_T;T]} (x',A' ; x, A) {\cal P}^{Cond}_t( x,A) 
  \bigg]
\label{markovcondjumpforward}
\end{eqnarray}
The difference with respect to the initial forward joint dynamics of Eq. \ref{markovjumpAforward}
is in the time-dependent forward rates  
\begin{eqnarray}
w_t^{Forw[x_T,A_T;T]} (x,A ; x', A') \equiv    P_{T,t}(x_T,A_T \vert x,A)  w(x,A;x',A')
   \frac{1}{P_{T,t}(x_T,A_T \vert x',A')} 
\label{wjumpforward}
\end{eqnarray}
that involve the conjugation of the joint rates $w(x,A;x',A')  \equiv  w(x;x') \delta(A-A'- \beta(x,x') )$
of Eq. \ref{wjoint} with the full propagators 
$P_{T-t-1}(x_T,A_T \vert x,A) $ and $P_{T,t}(x_T,A_T \vert x',A') $ up to the imposed final values $(x_T,A_T)$ at time $T$. Eq. \ref{wjumpforward} is the analog of Eq. \ref{markovcondforward} concerning discrete-time Markov chains.

(ii)  Backward perspective : Eq. \ref{markovcond} allows to replace all the propagators on $[t,T]$
\begin{eqnarray}
P_{T,t}( x_T,A_T \vert  x,A)=
 {\cal P}^{Cond}_t( x,A) \frac{P_{T,0}( x_T,A_T \vert  x_0,A_0)}{P_{t,0}( x,A \vert  x_0,A_0) }
\label{markovcondelimptb}
\end{eqnarray}
in Eq. \ref{markovcondjumpdericalcul} to obtain
\begin{eqnarray}
&& - \partial_t  {\cal P}^{Cond}_t(x,A) 
 =   
  \alpha(x)      \partial_{A}   {\cal P}^{Cond}_t( x,A)
\nonumber \\ &&
+   \sum_{x' \ne x } \int d A'  
\bigg[  {\cal P}^{Cond}_t( x',A') w_t^{Backw[x_0,A_0;0]} (x',A' ; x, A)
-  {\cal P}^{Cond}_t( x,A)     w_t^{Backw[x_0,A_0;0]} (x,A; x', A') \bigg]
\label{markovcondjumpbackward}
\end{eqnarray}
where the time-dependent backward rates  
\begin{eqnarray}
w_t^{Backw[x_0,A_0;0]} (x',A' ; x, A) 
\equiv   \frac{ 1 }{P_{t,0}(x',A' \vert x_0,A_0)}  w(x',A';x,A)  P_{t,0}(x,A \vert x_0,A_0) 
\label{wjumpbackward}
\end{eqnarray}
involve the conjugation of the joint rates $w(x,A;x',A')  \equiv  w(x;x') \delta(A-A'- \beta(x,x') )$
of Eq. \ref{wjoint} with the full propagators 
$P_{t,0}(x',A' \vert x_0,A_0)$ and $  P_{t,0}(x,A \vert x_0,A_0) $ up to the imposed initial values $(x_0,A_0)$ at time $t=0$.
Eq. \ref{wjumpbackward} is the analog of Eq. \ref{markovcondbackward} concerning discrete-time Markov chains.

(iii) The compatibility between the two dynamical equations of Eqs \ref{markovcondjumpforward}
 and \ref{markovcondjumpbackward} can be checked via their sum 
\begin{eqnarray}
&&0
  \stackrel{?}{=}    
  \sum_{x' \ne x } \int d A'  \bigg[ w_t^{Forw[x_T,A_T;T]} (x,A ; x', A')   {\cal P}^{Cond}_t( x',A') 
  - w_t^{Forw[x_T,A_T;T]} (x',A' ; x, A) {\cal P}^{Cond}_t( x,A) 
  \bigg]
\nonumber \\ &&
+   \sum_{x' \ne x } \int d A'  
\bigg[  {\cal P}^{Cond}_t( x',A') w_t^{Backw[x_0,A_0;0]} (x',A' ; x, A)
-  {\cal P}^{Cond}_t( x,A)     w_t^{Backw[x_0,A_0;0]} (x,A; x', A') \bigg]
\label{markovcondjumpcompatibility}
\end{eqnarray}
that is found to vanish using Eqs \ref{wjumpforward} \ref{wjumpbackward} and \ref{markovcond}.

(iv)  The half-difference between the two dynamical equations of 
Eqs \ref{markovcondjumpforward}
 and \ref{markovcondjumpbackward} 
 yields the new dynamical equation
\begin{eqnarray}
&& \partial_t  {\cal P}^{Cond}_t(x,A) 
 =    - \alpha(x)   \partial_A   {\cal P}^{Cond}_t( x,A)
\nonumber \\ &&
+  
  \sum_{x' \ne x } \int d A'  \bigg[ 
  w_t^{[x_T,A_T;T],[\vec x_0,A_0;0]} (x,A ; x', A')     {\cal P}^{Cond}_t( x',A') 
  -   w_t^{[x_T,A_T;T],[\vec x_0,A_0;0]} (x',A' ; x, A)  
   {\cal P}^{Cond}_t( x,A) 
  \bigg]
\label{markovcondjumpforwardbackward}
\end{eqnarray}
with the time-dependent rates
\begin{eqnarray}
&& w_t^{[x_T,A_T;T],[x_0,A_0;0]} (x,A ; x', A')  
 \equiv \frac{ w_t^{Forw[x_T,A_T;T]} (x,A ; x', A') -w_t^{Backw[x_0,A_0;0]} (x',A' ; x, A)  }{2}
\nonumber \\
&& =  \frac{1}{2} \bigg[  P_{T,t}(x_T,A_T \vert x,A)  w(x,A;x',A')   \frac{1}{P_{T,t}(x_T,A_T \vert x',A')} 
-   \frac{ 1 }{P_{t,0}(x',A' \vert x_0,A_0)}  w(x',A';x,A)  P_{t,0}(x,A \vert x_0,A_0) 
\bigg] \ \ \ 
\label{wjumptotal}
\end{eqnarray}


\subsection{ Simple example : conditioning the Sisyphus Markov Jump process on the number of resets}

Let us now consider the continuous-time analog of the Sisyphus Random Walk discussed in subsection \ref{subsec_resetchain}.
The Sisyphus Markov Jump process defined on the half-line $x=0,1,2,..$
is defined as follows :
when Sisyphus is at position $x$ at time $t$, he
can return to the origin $x=0$ with the reset rate $r$, he can move forward to the next position $(x+1)$ with rate $w$,
and otherwise he remains at its position $x$.

As time-additive observable $A$, we will choose the number of resets, so that 
the joint generator of Eq. \ref{wjoint} becomes
\begin{eqnarray}
w(x,A;x',A')  \equiv  r \delta_{x,0} \delta_{A,A'+1}   + w \delta_{x,x'+1}  \delta_{A,A'}  \ \  \ \ { \rm for } \ \ (x,A)\ne(x',A')
\label{wjointreset}
\end{eqnarray}

\subsubsection{ Joint propagator $P_{t,t_0}(x,A \vert x_0 A_0) $  }

The forward dynamics of Eq. \ref{markovjumpAforward} reads for the present model where $A$ is discrete
 with the generator of Eq. \ref{wjointreset}
\begin{eqnarray}
&& \partial _t P_{t,t_0}(x,A \vert x_0,A_0) 
 =  \sum_{(x',A') \ne (x,A) } \bigg[  w(x,A;x',A')  P_{t,t_0}(x',A' \vert x_0,A_0) -  w(x',A';x,A)P_{t,t_0}(x,A\vert x_0,A_0)\bigg]
\nonumber \\
&& =  r \delta_{x,0}  \theta( A>A_0)   \sum_{x'=0 }^{+\infty} P_{t,t_0}(x',A-1 \vert x_0,A_0) 
+ w \theta(x>0)      P_{t,t_0}(x-1,A \vert x_0,A_0) 
 - (r+w) P_{t,t_0}(x,A\vert x_0,A_0)
\label{markovjumpAforwardreset}
\end{eqnarray}
The solution can be directly written from the renewal analysis of the dynamics  
\begin{eqnarray}
&& P_{t,t_0}(x,A \vert x_0 A_0)  = e^{-r(t-t_0)} \delta_{A,A_0} \theta(x \geq x_0) \frac{[w (t-t_0) ]^{(x-x_0)} }{(t-t_0)!} e^{- w (t-t_0)}
\nonumber \\
&& + \theta( A>A_0) \theta(x \geq 0)  \frac{ r^{(A-A_0)} }{(A-A_0-1)!} e^{-r (t-t_0)} \frac{w^x}{x!}
\int_0^{(t-t_0)} d \tau [(t-t_0)-\tau]^{(A-A_0)-1} \tau^{x} e^{-w \tau}
\label{jointpropagatorsolutionjump}
\end{eqnarray}
The summation of Eq. \ref{jointpropagatorsolutionjump}
over the variable $x$ allows to recover that the propagator for the variable $A$ alone corresponds to the Poisson distribution for $(A-A_0)$ and is independent of $x_0$
\begin{eqnarray}
 \sum_{x=0}^{+\infty} P_{t,t_0}(x,A \vert x_0 A_0)  = 
  \theta( A\geq A_0) \frac{ [r (t-t_0)]^{(A-A_0)} }{(A-A_0)!} e^{-r (t-t_0)} 
 \equiv P_{t,t_0}(A \vert A_0)
\label{jointpropagatorsolutionAalonejump}
\end{eqnarray}
The summation of Eq. \ref{jointpropagatorsolutionjump}
over the variable $A$ yields the propagator for the initial Markov process $x(t)$ alone
\begin{eqnarray}
 \sum_{A=A_0}^{+\infty} P_{t,t_0}(x,A \vert x_0 A_0) && = 
 e^{-r(t-t_0)}  \theta(x \geq x_0) \frac{[w (t-t_0) ]^{(x-x_0)} }{(t-t_0)!} e^{- w (t-t_0)}
 +    \theta(x \geq 0)  r  \frac{w^x}{x!}
\int_0^{(t-t_0)} d \tau  \tau^{x} e^{-(w+r) \tau}
  \nonumber \\ 
  &&
  \equiv P_{t,t_0}(x \vert x_0)
\label{jointpropagatorsolutionxalonejump}
\end{eqnarray}
that converges towards the steady state corresponding to the geometric distribution
\begin{eqnarray}
P_{t,t_0}(x \vert x_0) \opsimeq_{(t-t_0) \to +\infty}     \theta(x \geq 0)  r  \frac{w^x}{x!}
\int_0^{+\infty} d \tau  \tau^{x} e^{-(w+r) \tau}
= \frac{r}{w+r} \left(  \frac{w}{w+r}\right)^x
 \equiv P_{st}(x )
\label{jointpropagatorsolutionxalonesteadyjump}
\end{eqnarray}


\subsubsection{ Forward generator of the conditioned dynamics  }

For the conditioned dynamics, the 
 forward generator of Eq. \ref{wjumpforward}
reads using the joint generator of Eq. \ref{wjointreset} for $(x,A)\ne(x',A') $
\begin{eqnarray}
 w_t^{Forw[x_T,A_T;T]} (x,A ; x', A') && =    P_{T,t}(x_T,A_T \vert x,A)  w(x,A;x',A')   \frac{1}{P_{T,t}(x_T,A_T \vert x',A')} 
\nonumber \\
&& = \delta_{x,0} \delta_{A,A'+1} r \frac{P_{T,t}(x_T,A_T \vert x,A'+1)  }{P_{T,t}(x_T,A_T \vert x',A')} 
  +  \delta_{x,x'+1}  \delta_{A,A'} w \frac{P_{T,t}(x_T,A_T \vert x'+1,A')  }{P_{T,t}(x_T,A_T \vert x',A')} 
  \nonumber \\
&& \equiv \delta_{x,0} \delta_{A,A'+1}  r^{Forw[x_T,A_T;T]}_{t}(x',A') 
  +  \delta_{x,x'+1}  \delta_{A,A'}   w^{Forw[x_T,A_T;T]}_{t}(x',A') 
\label{wjumpforwardreset}
\end{eqnarray}
So the conditioned dynamics corresponds to a Sisyphus Markov jump process, 
where the initial reset rate $r$ and the initial forward jump rate $w$
have been replaced by reset rates and forward jump rates that depend on the time $t$ and on the configuration $(x',A')$
\begin{eqnarray}
r^{Forw[x_T,A_T;T]}_{t}(x',A') && = r \frac{P_{T,t}(x_T,A_T \vert x,A'+1)  }{P_{T,t}(x_T,A_T \vert x',A')}
\nonumber \\
 w^{Forw[x_T,A_T;T]}_{t}(x',A') && =w \frac{P_{T,t}(x_T,A_T \vert x'+1,A')  }{P_{T,t}(x_T,A_T \vert x',A')}
\label{effectiveresetjump}
\end{eqnarray}
where on can plug the explicit form of the joint propagator given in Eq. \ref{jointpropagatorsolutionjump}.

The 'canonical conditioning' (see the reminder in the two Appendices) of the Sisyphus Markov jump process 
has been studied in \cite{c_reset} for the more general case where the reset rates of the initial model are space-dependent $r_x$ (instead of being given by the constant value $r$)
and where the time-additive observable involve two arbitrary functions $\alpha(x)$ and $\beta(x,x')$.


\section{ Application to diffusion processes in dimension $d$ }

\label{sec_diff}

In this section, we consider the diffusion process $\vec x(t)$, where 
the $d$ components $x_i(t)$ for $i=1,..,d$ follow the Langevin 
stochastic differential equations involving the functions $(f_i[\vec x],g_i[\vec x])$
and $d$ independent Brownian motions $B_i(t)$ 
\begin{eqnarray}
dx_i(t) =  f_i[ \vec x (t) ] \ dt + g_i [\vec x (t) ]  \ dB_i(t)
\label{langevin}
\end{eqnarray}
in the Stratonovich interpretation.
Equivalently, the dynamics can be defined via
the Fokker-Planck equation for the probability $P_t(\vec x) $ to be at position $\vec x$ at time $t$
\begin{eqnarray}
 \partial_t P_t(\vec x)    =    \sum_{i=1}^d \partial_{x_i}  
   \bigg[ - F_i[ \vec x ]   P_t(\vec x ) + D_i [ \vec x ]  \partial_{x_i} P_t(\vec x) \bigg]
\label{fokkerplanck}
\end{eqnarray}
with the following components for the force and for the diffusion coefficient 
\begin{eqnarray}
F_i[ \vec x ]    &&  =   f_i[ \vec x ] -  \frac{g_i [\vec x] \partial_{x_i} g_i [\vec x]}{2}  
\nonumber \\
D_i [ \vec x ] && =  \frac{g_i^2 [\vec x]}{2}  
\label{fokkerplancklangevin}
\end{eqnarray}

The time-additive observable $A(t)$ of the trajectory $\vec x(t_1 \leq s \leq t_2)$ of Eq. \ref{additivetraj}
can be parametrized by the function $\alpha[\vec x]$ and by the field $\vec \beta[\vec x]$
in the Stratonovich interpretation
\begin{eqnarray}
A(t_2)-A(t_1)={\cal A}[x(t_1 \leq s \leq t_2) ] =    \int_{t_1}^{t_2} \left[   \alpha[\vec x(s)] ds + \vec\beta[ \vec x(s)] .  d \vec x(s) \right]
\label{additivediff}
\end{eqnarray}

\subsection{  Dynamics of the joint propagator $P_{t,t_0}(\vec x,A \vert \vec x_0,A_0)$  }

Since the increment between $t$ and $(t+dt)$ of the time-additive observable $A(t)$ of Eq. \ref{additivediff}
can be rewritten in terms of the $d$ Langevin increments $ d  x_i(t)$ of Eq. \ref{langevin}
\begin{eqnarray}
dA(t) && = A(t+dt)-A(t)  =   \alpha[\vec x(t)]  dt +  \sum_{i=1}^d  \beta_i [ \vec x(t)] . d  x_i(t)
\nonumber \\
&& =  \left[  \alpha[\vec x(t)]  +\sum_{i=1}^d   \beta_i [ \vec x(t)] f_i[ \vec x (t) ] \right] dt
 +   \sum_{i=1}^d   \beta_i [ \vec x(t)]   g_i [\vec x (t) ]  dB_i(t) 
\label{additivediffelementary}
\end{eqnarray}
one can consider that $A(t)$ is a supplementary $(d+1)$ coordinate
 for the Langevin system in the Stratonovich interpretation
 of Eq. \ref{langevin}, that involves the $d$ previous Brownian motions $B_i(t)$.
As a consequence,  one can write the Fokker-Planck equations generalizing Eq. \ref{fokkerplanck}
as follows : 

(i) the forward generator
\begin{eqnarray}
 {\cal F}  
 =  - \alpha[\vec x] \partial_A  
   -  \sum_{i=1}^d \left( \partial_{x_i}  +  \beta_i [ \vec x] \partial_A \right)   F_i[ \vec x ]  
   +   \sum_{i=1}^d   \left( \partial_{x_i}  +  \beta_i [ \vec x] \partial_A \right) D_i [\vec x]  
 \left( \partial_{x_i}  +  \beta_i [ \vec x] \partial_A \right) 
 \label{fokkerplanckAgenerator}
\end{eqnarray}
governs the forward Fokker-Planck equation
of the joint propagator $P_{t,0}(\vec x,A \vert \vec x_0,A_0) $  with respect to the final variables $(\vec x,A) $ 
at time $t$
\begin{eqnarray}
  \partial_t P_{t,0}(\vec x,A \vert \vec x_0,A_0) && = {\cal F} P_{t,0}(\vec x,A \vert \vec x_0,A_0)  
 \nonumber \\
 &&= - \alpha[\vec x] \partial_A P_{t,0}(\vec x,A \vert \vec x_0,A_0)  
  -  \sum_{i=1}^d \left( \partial_{x_i}  +  \beta_i [ \vec x] \partial_A \right)   \bigg[  F_i[ \vec x ]   P_{t,0}(\vec x,A \vert \vec x_0,A_0) \bigg]
 \nonumber \\
 &&  +  \sum_{i=1}^d \left( \partial_{x_i}  +  \beta_i [ \vec x] \partial_A \right)   \bigg[  D_i[\vec x] 
 \left( \partial_{x_i}  +  \beta_i [ \vec x] \partial_A \right)
    P_{t,0}(\vec x,A \vert \vec x_0,A_0) \bigg]
\label{fokkerplanckAforward}
\end{eqnarray}

(ii) the backward generator corresponding to the adjoint differential operator of Eq. \ref{fokkerplanckAgenerator}
 \begin{eqnarray}
 {\cal F}^{\dagger} 
 =   \alpha[\vec x] \partial_A  
   +   \sum_{i=1}^d F_i[ \vec x ]  \left( \partial_{x_i}  +  \beta_i [ \vec x] \partial_A \right)   
   +      \sum_{i=1}^d   \left( \partial_{x_i}  +  \beta_i [ \vec x] \partial_A \right) D_i [\vec x]  
 \left( \partial_{x_i}  +  \beta_i [ \vec x] \partial_A \right) 
 \label{fokkerplanckAgeneratordagger}
\end{eqnarray}
governs the backward Fokker-Planck equation for the joint propagator $P_{T,t}(\vec x_T,A_T \vert \vec x,A) $ 
 with respect to the initial variables $(\vec x,A)$ at time $t$
\begin{eqnarray}
 - \partial_t P_{T,t}(\vec x_T,A_T \vert \vec x,A) && = {\cal F}^{\dagger } P_{T,t}(\vec x_T,A_T \vert \vec x,A)
 \nonumber \\
&& =   \alpha[\vec x] \partial_A  P_{T,t}(\vec x_T,A_T \vert \vec x,A)
   +  \sum_{i=1}^d F_i[ \vec x ]  
 \left( \partial_{x_i}  +  \beta_i [ \vec x] \partial_A \right)   P_{T,t}(\vec x_T,A_T \vert \vec x,A) 
 \nonumber \\
&& 
 +  \sum_{i=1}^d \left( \partial_{x_i}  +  \beta_i [ \vec x] \partial_A \right)   \bigg[  D_i[\vec x] 
 \left( \partial_{x_i}  +  \beta_i [ \vec x] \partial_A \right)
    P_{T,t}(\vec x_T,A_T \vert \vec x,A)   \bigg]    
\label{fokkerplanckAbackward}
\end{eqnarray}


\subsection{ Markov dynamics for the conditional probability ${\cal P}^{Cond}_t(\vec x,A) $ with time-dependent additional forces }

Let us now focus on the dynamics for the conditional probability of Eq. \ref{markovcond}
\begin{eqnarray}
{\cal P}^{Cond}_t(\vec x,A) =  \frac{P_{T,t}(\vec x_T,A_T \vert \vec x,A) P_{t,0}(\vec x,A \vert \vec x_0,A_0)}
{P_{T,0}(\vec x_T,A_T \vert \vec x_0,A_0)}
\label{markovconddiff}
\end{eqnarray}
Its dynamics with respect to the time $t$
involves 
the forward dynamics with generator ${\cal F}$ of Eq. \ref{fokkerplanckAforward} for the propagator $P_{t,0}(\vec x,A \vert \vec x_0,A_0) $ 
and the backward dynamics with generator ${\cal F}^{\dagger} $ of Eq. \ref{fokkerplanckAbackward}
for the propagator $P_{T,t}(\vec x_T,A_T \vert \vec x,A) $.
So the time-derivative of the conditional probability of Eq. \ref{markovconddiff}
reads
\begin{eqnarray}
\partial_t  {\cal P}^{Cond}_t(\vec x,A) && = \left[ \partial_t P_{T,t}(\vec x_T,A_T \vert \vec x,A)\right] 
\frac{ P_{t,0}(\vec x,A \vert \vec x_0,A_0)}{P_{T,0}(\vec x_T,A_T \vert \vec x_0,A_0)}
+  \frac{P_{T,t}(\vec x_T,A_T \vert \vec x,A) }{P_{T,0}(\vec x_T,A_T \vert \vec x_0,A_0)} 
\left[ \partial_t P_{t,0}(\vec x,A \vert \vec x_0,A_0) \right]
\nonumber \\
&& =- \frac{ P_{t,0}(\vec x,A \vert \vec x_0,A_0)}{P_{T,0}(\vec x_T,A_T \vert \vec x_0,A_0)}
\left[  {\cal F}^{\dagger} P_{T,t}(\vec x_T,A_T \vert \vec x,A)\right] 
+  \frac{P_{T,t}(\vec x_T,A_T \vert \vec x,A) }{P_{T,0}(\vec x_T,A_T \vert \vec x_0,A_0)} 
\left[ {\cal F} P_{t,0}(\vec x,A \vert \vec x_0,A_0) \right]
\label{markovconddiffderi}
\end{eqnarray}

(i) Forward perspective : Eq. \ref{markovconddiff} allows to plug the propagator
\begin{eqnarray}
P_{t,0}(\vec x,A \vert \vec x_0,A_0) = {\cal P}^{Cond}_t(\vec x,A) 
\frac{P_{T,0}(\vec x_T,A_T \vert \vec x_0,A_0)}{P_{T,t}(\vec x_T,A_T \vert \vec x,A)}
\label{markovconddiffelimpt}
\end{eqnarray}
into Eq. \ref{markovconddiffderi} to obtain
\begin{eqnarray}
\partial_t  {\cal P}^{Cond}_t(\vec x,A) 
&& =- \frac{  {\cal P}^{Cond}_t(\vec x,A) }{P_{T,t}(\vec x_T,A_T \vert \vec x,A)}
\left[  {\cal F}^{\dagger} P_{T,t}(\vec x_T,A_T \vert \vec x,A)\right] 
 + P_{T,t}(\vec x_T,A_T \vert \vec x,A)\left[ {\cal F}  
\frac{{\cal P}^{Cond}_t(\vec x,A)  }{P_{T,t}(\vec x_T,A_T \vert \vec x,A)} \right]
\label{markovconddiffderiforwardcalcul}
\end{eqnarray}
or more explicitly using the forms of Eqs \ref{fokkerplanckAgenerator} and \ref{fokkerplanckAgeneratordagger}
for the differential generator ${\cal F}$ and its adjoint ${\cal F}^{\dagger}$
\begin{eqnarray}
 \partial_t  {\cal P}^{Cond}_t(\vec x,A) 
&& =
  -  \alpha[\vec x]   \partial_A  {\cal P}^{Cond}_t(\vec x,A)  
    -  \sum_{i=1}^d 
\left( \partial_{x_i}  +  \beta_i [ \vec x] \partial_A \right) 
\bigg[  \big( F_i[ \vec x ]  + F^{Forw[\vec x_T,A_T;T]}_i[ \vec x,A ; t ]\big){\cal P}^{Cond}_t(\vec x,A) \bigg]
 \nonumber \\ &&  
   +  \sum_{i=1}^d   
\left( \partial_{x_i}  +  \beta_i [ \vec x] \partial_A \right)
\left[ D_i [\vec x]  
 \left( \partial_{x_i}  +  \beta_i [ \vec x] \partial_A \right)
  {\cal P}^{Cond}_t(\vec x,A)     \right] 
\label{markovconddiffderiforward}
\end{eqnarray}
where the only differences with respect to the forward joint Fokker-Planck dynamics of Eq. \ref{fokkerplanckAforward}
are the additional time-dependent forces 
\begin{eqnarray}
  F^{Forw[\vec x_T,A_T;T]}_i[ \vec x,A ; t ] \equiv 
 2 D_i [\vec x] \left( \partial_{x_i}  +  \beta_i [ \vec x] \partial_A \right)  \ln P_{T,t}(\vec x_T,A_T \vert \vec x,A)
\label{forcesupforward}
\end{eqnarray}
that involve the propagator $P_{T,t}(\vec x_T,A_T \vert \vec x,A) $
 up to the imposed final values $(\vec x_T,A_T)$ at time $T$.
Eq. \ref{forcesupforward} is the analog of
Eqs \ref{markovcondforward} and \ref{wjumpforward}.

(ii) Backward perspective : Eq. \ref{markovconddiff} allows to plug the propagator
\begin{eqnarray}
P_{T,t}(\vec x_T,A_T \vert \vec x,A)  = {\cal P}^{Cond}_t(\vec x,A) 
\frac{P_{T,0}(\vec x_T,A_T \vert \vec x_0,A_0)}{P_{t,0}(\vec x,A \vert \vec x_0,A_0) }
\label{markovconddiffelimptt}
\end{eqnarray}
into Eq. \ref{markovconddiffderi} to obtain
\begin{eqnarray}
- \partial_t  {\cal P}^{Cond}_t(\vec x,A) 
 = P_{t,0}(\vec x,A \vert \vec x_0,A_0)
\left[  {\cal F}^{\dagger}  \frac{{\cal P}^{Cond}_t(\vec x,A)  }{P_{t,0}(\vec x,A \vert \vec x_0,A_0) }\right] 
-   
\frac{ {\cal P}^{Cond}_t(\vec x,A)}{P_{t,0}(\vec x,A \vert \vec x_0,A_0) }
\left[ {\cal F} P_{t,0}(\vec x,A \vert \vec x_0,A_0) \right]
\label{markovconddiffderibackward}
\end{eqnarray}
or more explicitly using the forms of Eqs \ref{fokkerplanckAgenerator} and \ref{fokkerplanckAgeneratordagger}
for the differential generator ${\cal F}$ and its adjoint ${\cal F}^{\dagger}$
\begin{eqnarray}
-  \partial_t  {\cal P}^{Cond}_t(\vec x,A) 
&& =
    \alpha[\vec x]   \partial_A  {\cal P}^{Cond}_t(\vec x,A)  
    +  \sum_{i=1}^d 
\left( \partial_{x_i}  +  \beta_i [ \vec x] \partial_A \right) 
\bigg[  \big( F_i[ \vec x ]  + F^{Backw[\vec x_0,A_0;0]}_i[ \vec x,A; t ]\big){\cal P}^{Cond}_t(\vec x,A) \bigg]
 \nonumber \\ &&  
  +   \sum_{i=1}^d   
\left( \partial_{x_i}  +  \beta_i [ \vec x] \partial_A \right)
\bigg[ D_i [\vec x]  
 \left( \partial_{x_i}  +  \beta_i [ \vec x] \partial_A \right)
  {\cal P}^{Cond}_t(\vec x,A)     \bigg]
\label{dyncondback}
\end{eqnarray}
where the time-dependent forces 
\begin{eqnarray}
 F^{Backw[\vec x_0,A_0;0]}_i[ \vec x,A; t ] \equiv 
 -  2 D_i [\vec x] \left( \partial_{x_i}  +  \beta_i [ \vec x] \partial_A \right)  \ln P_{t,0}(\vec x,A \vert \vec x_0,A_0)
\label{forcesupbackward}
\end{eqnarray}
involve the propagator $ P_{t,0}(\vec x,A \vert \vec x_0,A_0)$
up to the imposed initial values $(\vec x_0,A_0)$ at time $t=0$
Eq. \ref{forcesupbackward} is the analog of Eqs \ref{markovcondbackward} and \ref{wjumpbackward}.

(iii) The compatibility between the two dynamical equations of Eqs \ref{markovconddiffderiforward}
 and \ref{markovconddiffderibackward} can be checked via their sum that 
 can be evaluated using the explicit expressions of Eqs \ref{forcesupforward}
 and \ref{forcesupbackward}
 for the additional forces
\begin{small}
\begin{eqnarray}
0   
&& \stackrel{?}{=} 
 \sum_{i=1}^d 
\left( \partial_{x_i}  +  \beta_i [ \vec x] \partial_A \right) 
\bigg[  \big( F^{Backw[\vec x_0,A_0;0]}_i[ \vec x,A; t ]  - F^{Forw[\vec x_T,A_T;T]}_i[ \vec x,A ; t ]\big){\cal P}^{Cond}_t(\vec x,A) +2 D_i [\vec x]  
 \left( \partial_{x_i}  +  \beta_i [ \vec x] \partial_A \right)
  {\cal P}^{Cond}_t(\vec x,A)      \bigg]
 \nonumber \\
&& = \sum_{i=1}^d 
\left( \partial_{x_i}  +  \beta_i [ \vec x] \partial_A \right) 2 D_i [\vec x] {\cal P}^{Cond}_t(\vec x,A)
\bigg[  
 \left( \partial_{x_i}  +  \beta_i [ \vec x] \partial_A \right)
  \ln \frac{ {\cal P}^{Cond}_t(\vec x,A)   }{ P_{T,t}(\vec x_T,A_T \vert \vec x,A)P_{t,0}(\vec x,A \vert \vec x_0,A_0) }    \bigg] 
  \nonumber \\
&& = \sum_{i=1}^d 
\left( \partial_{x_i}  +  \beta_i [ \vec x] \partial_A \right) 2 D_i [\vec x] {\cal P}^{Cond}_t(\vec x,A)
\bigg[  
 \left( \partial_{x_i}  +  \beta_i [ \vec x] \partial_A \right)
  \ln \frac{ 1   }{ P_{T,0}(\vec x_T,A_T  \vert \vec x_0,A_0) }    \bigg]  
  =0
\label{markovcheck}
\end{eqnarray}
\end{small}

where we have used Eq. \ref{markovconddiff} to obtain the propagator $P_{T,0}(\vec x_T,A_T  \vert \vec x_0,A_0) $ that does not depend upon $(\vec x,A)$.

(iv) The half-difference between the two dynamical equations of Eqs \ref{markovconddiffderiforward}
 and \ref{markovconddiffderibackward}
 leads to the new dynamical equation involving only drift contributions
\begin{eqnarray}
 \partial_t  {\cal P}^{Cond}_t(\vec x,A) 
 =
  -  \alpha[\vec x]   \partial_A  {\cal P}^{Cond}_t(\vec x,A)  
    -  \sum_{i=1}^d 
\left( \partial_{x_i}  +  \beta_i [ \vec x] \partial_A \right) 
\bigg[  \big( F_i[ \vec x ]  +  F^{[\vec x_T,A_T;T],[\vec x_0,A_0;0]}_i[ \vec x,A ; t]  \big){\cal P}^{Cond}_t(\vec x,A) \bigg] 
\label{markovconddiffderiforwardbackward}
\end{eqnarray}
where the time-dependent additional forces
\begin{eqnarray}
 F^{[\vec x_T,A_T;T],[\vec x_0,A_0;0]}_i[ \vec x,A ; t] 
 && \equiv \frac{F^{Forw[\vec x_T,A_T;T]}_i[ \vec x,A ; t ]+ F^{Backw[\vec x_0,A_0;0]}_i[ \vec x,A; t ] }{2}
\nonumber \\
&&   = D_i [\vec x] \left( \partial_{x_i}  +  \beta_i [ \vec x] \partial_A \right)  \ln \frac{ P_{T,t}(\vec x_T,A_T \vert \vec x,A) }
   { P_{t,0}(\vec x,A \vert \vec x_0,A_0) }
\label{forcesupforwardav}
\end{eqnarray}
involves both propagators $ P_{t,0}(\vec x,A \vert \vec x_0,A_0)$ and $P_{T,t}(\vec x_T,A_T \vert \vec x,A) $.


\subsection{  Stratonovich stochastic differential equations for the conditioned process $(\vec x^*(t),A^*(t))$}

The forward Fokker-Planck dynamics of Eq. \ref{markovconddiffderiforward}
can be translated into 
the following Stratonovich stochastic differential equations for the joint conditioned process $(\vec x^*(t),A^*(t))$.
The $d$ components $x_i^*(t)$ for $i=1,..,d$ of $\vec x^*(t) $ satisfy the Stratonovich 
stochastic differential equations in terms of $d$ independent Brownian motions $B_i(t)$ 
\begin{eqnarray}
dx_i^*(t) =  \left(f_i[ \vec x^* (t) ] + F^{Forw[\vec x_T,A_T;T]}_i[ \vec x^*(t),A^*(t) ; t ] \right) dt + g_i [\vec x^* (t) ]  \ dB_i(t)
\label{langevinstar}
\end{eqnarray}
where the only differences with respect to the unconditioned case of Eq. \ref{langevin}
are the additional time-dependent forces $F^{Forw[\vec x_T,A_T;T]}_i[ \vec x^*,A^* ; t ] $ given in Eq. \ref{forcesupforward}.

Since the increment between $t$ and $(t+dt)$ of the time-additive observable $A^*(t)$ 
can be rewritten in terms of $\vec x^*(t) $ and of the $d$ Langevin increments $ d  x_i^*(t)$ of Eq. \ref{langevinstar},
the Stratonovich 
stochastic differential equation for $A^*(t) $ reads
\begin{eqnarray}
&& dA^*(t)  = A^*(t+dt)-A^*(t)  =   \alpha[\vec x^*(t)]  dt +  \sum_{i=1}^d  \beta_i [ \vec x^*(t)] . d  x_i^*(t)
\nonumber \\
&& =  \left[  \alpha[\vec x^*(t)]  +\sum_{i=1}^d   \beta_i [ \vec x^*(t)]  \left(f_i[ \vec x^* (t) ] 
+ F^{Forw[\vec x_T,A_T;T]}_i[ \vec x^*(t),A^*(t) ; t ] \right)\right] dt
 +   \sum_{i=1}^d   \beta_i [ \vec x^*(t)]   g_i [\vec x^* (t) ]  dB_i(t) 
 \ \ \ 
\label{additivestar}
\end{eqnarray}
The Stratonovich Stochastic Differential Equations of Eqs \ref{langevinstar}
and \ref{additivestar}
can be then used to generate stochastic trajectories of the conditioned process $(\vec x^*(t),A^*(t))$.


\subsection{ Simple example :  Brownian $B(t)$ as time-additive observable of the diffusion process $x(t)$}

\subsubsection{ Brownian motion $B(t)$ conditioned on the value of the diffusion process $x(t)$ }

Let us consider the one-dimensional diffusion process of Eq. \ref{langevin}
\begin{eqnarray}
dx(t) =  f[  x (t) ] \ dt + g [ x (t) ]  \ dB(t)
\label{langevin1d}
\end{eqnarray}
associated to the Fokker-Planck Eq. \ref{fokkerplanck}
\begin{eqnarray}
 \partial_t P_t( x)    =     \partial_{x}  
   \bigg[ - F[  x ]   P_t( x ) + D [  x ]  \partial_{x} P_t( x) \bigg]
\label{fokkerplanck1d}
\end{eqnarray}
with the the force and the diffusion coefficient of Eq. \ref{fokkerplancklangevin}
\begin{eqnarray}
F[ x ]    &&  =   f[  x ] -  \frac{g [ x] g' [ x]}{2}  
\nonumber \\
D [  x ] && =  \frac{g^2 [ x]}{2}  
\label{fokkerplancklangevin1d}
\end{eqnarray}

As time-additive observable, let us choose the Brownian motion $B(t)$ satisfying Eq. \ref{langevin1d}
\begin{eqnarray}
dB(t) = - \frac{    f[  x (t) ] }{ g [ x (t) ] } dt + \frac{ 1 }{ g [ x (t) ] }  dx(t) 
\label{langevin1dforB}
\end{eqnarray}
i.e. the two functions $\alpha$ and $\beta$ are present in the parametrization of Eq. \ref{additivediff}
\begin{eqnarray}
 \alpha[ x]  && =- \frac{    f[  x  ] }{ g [ x  ] }
 \nonumber \\
  \beta[  x] && = \frac{ 1 }{ g [ x  ] }
\label{additivediffab}
\end{eqnarray}

The joint propagator $P_{t,t_0}( x,B \vert  x_0,B_0) $  
satisfies the forward dynamics of Eq. \ref{fokkerplanckAforward}
\begin{eqnarray}
  \partial_t P_{t,t_0}( x,B \vert  x_0,B_0)  && = 
    \partial_x \left[ -F[ x] P_{t,t_0}( x,B \vert  x_0,B_0) + D[ x] \partial_x P_{t,t_0}( x,B \vert  x_0,B_0)
  + \sqrt{ 2 D[ x] } \partial_B P_{t,t_0}( x,B \vert  x_0,B_0)  \right]
\nonumber \\
&&   
+  \frac{1}{2} \partial_B^2 P_{t,t_0}( x,B \vert  x_0,B_0) 
\label{fokkerplanckAforwardB}
\end{eqnarray}

Since in the microcanonical conditioning framework one considers the joint process $(x(t),B(t))$,
one can rephrase the 'conditioning of the diffusion process $x(t)$ on its time-additive observable $B(t)$' described above
as the 'conditioning of Brownian motion $B(t)$ on the diffusion process $x(t)$'.
This rephrasing is interesting because the diffusion process $x(t)$ generated via Eq. \ref{langevin1d}
is not a time-additive observable of the Brownian motion $B(t)$.

When the joint propagator $ P_{t,t_0}( x,B \vert  x_0,B_0)$ of Eq. \ref{fokkerplanckAforwardB}
is explicit, one can compute the
 additional time-dependent force of Eq. \ref{forcesupforward}
\begin{eqnarray}
  F^{Forw[ x_T,B_T;T]}[  x,B ; t ] = 
 2 D [ x] \left( \partial_x  +  \beta [  x] \partial_B \right)  \ln P_{T,t}( x_T,B_T \vert  x,B)
\label{forcesupforward1d}
\end{eqnarray}
that appear in the Stratonovich stochastic differential equations for the conditioned process $(x^*(t),B^*(t))$ 
as follows.
The Stratonovich stochastic differential equation
of Eq \ref{langevinstar} for $x^*(t)$ involves a Wiener process $W(t)$
\begin{eqnarray}
dx^*(t) =  \left(f[  x^* (t) ] + F^{Forw[\vec x_T,B_T;T]}[ x^*(t),B^*(t) ; t ] \right) dt + g [ x^* (t) ]  \ dW(t)
\label{langevinstar1d}
\end{eqnarray}
while the Stratonovich stochastic differential equation of Eq. \ref{additivestar} for $B^*(t) $ reads
using Eqs \ref{langevin1dforB} and \ref{additivediffab}
\begin{eqnarray}
dB^*(t) && =      \frac{ \left[ -  f[  x^* (t) ] dt +   dx^*(t) \right] }{ g [ x^* (t) ]  }  
\nonumber \\
&& =   \frac{   F^{Forw[\vec x_T,B_T;T]}[ x^*(t),B^*(t) ; t ]    }{ g [ x^* (t) ]  } dt  +dW(t)
\label{additivestar1d}
\end{eqnarray}
The dynamics of 
Eq. \ref{langevinstar1d} can also be rewritten in terms of the increment $dB^*(t) $ of Eq. \ref{additivestar1d} as
\begin{eqnarray}
dx^*(t) = f[  x^* (t) ]  dt + g [ x^* (t) ]  dB^*(t)
\label{langevinstar1dsimpler}
\end{eqnarray}


\subsubsection{ Explicit solution when $x(t)$ is the Ornstein-Uhlenbeck process }

In order to have a simple Gaussian solution for the joint propagator, let us now consider the case of the Ornstein-Uhlenbeck process for $x(t)$
corresponding to the constant diffusion coefficient and the linear restoring force 
\begin{eqnarray}
D [  x ] && =  D
\nonumber \\
F[ x ]    &&  =  -x 
\label{OUdef}
\end{eqnarray}
so that the corresponding stochastic differential equation reads 
\begin{eqnarray}
dx(t) =  - x (t) dt + \sqrt{2D}  \ dB(t)
\label{langevin1dOU}
\end{eqnarray}
both in the Stratonovich and in the Ito interpretations, since the diffusion constant is space-independent.

The forward Fokker-planck Eq. \ref{fokkerplanckAforwardB} for the joint propagator $P_{t,t_0}( x,B \vert  x_0,B_0) $
\begin{eqnarray}
  \partial_t P_{t,t_0}( x,B \vert  x_0,B_0)  && = 
    \partial_x \left[ x P_{t,t_0}( x,B \vert  x_0,B_0) \right] + D \partial^2_x P_{t,t_0}( x,B \vert  x_0,B_0)
  + \sqrt{ 2 D } \partial_B \partial_x P_{t,t_0}( x,B \vert  x_0,B_0)  
\nonumber \\
&&   
+  \frac{1}{2} \partial_B^2 P_{t,t_0}( x,B \vert  x_0,B_0) 
\label{fokkerplanckAforwardBou}
\end{eqnarray}
can be translated via the double Fourier transform
\begin{eqnarray}
 {\hat P}_{t,t_0}( q,k \vert  x_0,B_0) \equiv \int_{-\infty}^{+\infty} dx e^{iqx} \int_{-\infty}^{+\infty} dB e^{ikB} P_{t,t_0}( x,B \vert  x_0,B_0)
\label{fourier}
\end{eqnarray}
into the dynamical equation
\begin{eqnarray}
  \partial_t  {\hat P}_{t,t_0}( q,k \vert  x_0,B_0)  && = 
    -q \partial_q  {\hat P}_{t,t_0}( q,k \vert  x_0,B_0)
    - \left[ D q^2 + \sqrt{ 2 D } q k +  \frac{k^2}{2} \right]  {\hat P}_{t,t_0}( q,k \vert  x_0,B_0)
\label{fokkerplanckAforwardBoufourier}
\end{eqnarray}
and the initial conditions at $t=t_0$
\begin{eqnarray}
 {\hat P}_{t=t_0,t_0}( q,k \vert  x_0,B_0) \equiv \int_{-\infty}^{+\infty} dx e^{iqx} \int_{-\infty}^{+\infty} dB e^{ikB} 
 \delta(x-x_0) \delta(B-B_0) = e^{iqx_0}  e^{ikB_0} 
 \label{fourierini}
\end{eqnarray}
The solution 
\begin{eqnarray}
 {\hat P}_{t,t_0}( q,k \vert  x_0,B_0)   = e^{ -  q^2 \frac{D}{2} \left[ 1- e^{- 2(t-t_0)}\right]
 -  k^2  \frac{t-t_0}{2}   -  kq \sqrt{ 2 D }  \left[ 1- e^{- (t-t_0)}\right] + i  q x_0 e^{-(t-t_0) }+ i  k B_0 }
\label{gauss}
\end{eqnarray}
corresponds via the double-inverse Fourier transform of Eq. \ref{fourier}
to the bivariate Gaussian distribution
\begin{eqnarray}
&& P_{t,t_0}( x,B \vert  x_0,B_0)
 =  \int_{-\infty}^{+\infty} \frac{dq}{2\pi} e^{-iqx} \int_{-\infty}^{+\infty} \frac{dk}{2\pi}  e^{-ikB}  {\hat P}_{t,t_0}( q,k \vert  x_0,B_0)
\nonumber \\
  &&  =   
   \frac{1}{ 2 \pi \sigma(t,t_0) v(t,t_0) \sqrt{1-c^2(t,t_0)} } 
  e^{  - \frac{1}{2 [1-c^2(t,t_0)] } 
\left[   \left( \frac{ x- x_0 e^{-(t-t_0)} }{\sigma(t,t_0)}\right)^2
+\left(\frac{ B- B_0 }{v(t,t_0)} \right)^2
  - 2 c(t,t_0) \left( \frac{ x- x_0 e^{-(t-t_0)} }{\sigma(t,t_0)}\right)\left(\frac{ B- B_0 }{v(t,t_0)} \right)
   \right]    } 
\label{gaussinverse}
\end{eqnarray}
with the two variances
\begin{eqnarray}
\sigma^2(t,t_0)      && =   D \left[ 1- e^{- 2(t-t_0)}\right]
     \nonumber \\
v^2(t,t_0)       && =     ( t-t_0 )
\label{var}
\end{eqnarray}
and the rescaled correlation
\begin{eqnarray}
c(t,t_0) = \sqrt{  \frac{  2    \left[ 1- e^{- (t-t_0)}\right]}{  (t-t_0) \left[ 1+ e^{- (t-t_0)}\right] } }
\label{correl}
\end{eqnarray}

The conditioned forward dynamics is then governed by the Fokker-Planck Eq. \ref{markovconddiffderiforward}
\begin{eqnarray}
 \partial_t  {\cal P}^{Cond}_t( x,B) 
&& =
  - \partial_x \left[ \left( - x +  F^{Forw[ x_T,B_T;T]}_t[ x ,B  ]  \right) P_{t,t_0}( x,B \vert  x_0,B_0) \right] 
\nonumber \\
&&  + D \partial^2_x P_{t,t_0}( x,B \vert  x_0,B_0)
  + \sqrt{ 2 D } \partial_B \partial_x P_{t,t_0}( x,B \vert  x_0,B_0)     
+  \frac{1}{2} \partial_B^2 P_{t,t_0}( x,B \vert  x_0,B_0) 
 \label{markovconddiffderiforwardOU}
\end{eqnarray}
where the only difference with respect to the forward Fokker-Planck dynamics of Eq. \ref{fokkerplanckAforwardBou}
is the additional time-dependent force of Eq. \ref{forcesupforward}
\begin{eqnarray}
  F^{Forw[ x_T,B_T;T]}_t[ x ,B  ] \equiv 
 2 D \left( \partial_{x}  +  \frac{1}{\sqrt{2D} } \partial_B \right)  \ln P_{T,t}( x_T,B_T \vert  x,B)
\label{forcesupforwardOU}
\end{eqnarray}
Using the explicit form of Eq. \ref{gaussinverse}
for the propagator, one obtains 
\begin{eqnarray}
&& \ln P_{T,t}( x_T,B_T \vert  x,B)   =   
   -\ln \left( 2 \pi \sigma(T,t) v(T,t) \sqrt{1-c^2(T,t)}  \right)
  \nonumber \\
  &&   - \frac{1}{2 [1-c^2(T,t)] } 
\left[   \left( \frac{  x e^{-(T-t)} -x_T }{\sigma(T,t)}\right)^2
+\left(\frac{  B -B_T }{v(T,t)} \right)^2
  - 2 c(T,t) \left( \frac{  x e^{-(T-t)} -x_T}{\sigma(T,t)}\right)\left(\frac{  B -B_T}{v(T,t)} \right)
   \right]     
\label{gaussinverselast}
\end{eqnarray}
with the corresponding partial derivatives with respect to $x$ 
\begin{eqnarray}
 \partial_{x}  \ln P_{T,t}( x_T,B_T \vert  x,B)   =  
  \frac{e^{-(T-t)}}{\sigma(T,t) [1-c^2(T,t)] } 
\left[  -  \left(  \frac{  x e^{-(T-t)} -x_T }{\sigma(T,t)} \right)
  +    c(T,t) \left(\frac{  B -B_T}{v(T,t)} \right)
   \right]       
\label{gaussinverselastderix}
\end{eqnarray}
and with respect to $B$
\begin{eqnarray}
 \partial_{B} \ln P_{T,t}( x_T,B_T \vert  x,B)   =   
  \frac{1}{v(T,t) [1-c^2(T,t)] } 
\left[ - \left( \frac{  B -B_T }{v(T,t)} \right)
  +  c(T,t) \left( \frac{  x e^{-(T-t)} -x_T}{\sigma(T,t)}\right) 
   \right]      
\label{gaussinverselastderiB}
\end{eqnarray}
As a consequence, the additional time-dependent force of Eq. \ref{forcesupforwardOU} is linear with respect to $x$ and with respect to $B$
\begin{eqnarray}
  F^{Forw[ x_T,B_T;T]}_t[ x ,B  ]  
&& = \left[ \sqrt{2D}  \partial_{x}   \ln P_{T,t}( x_T,B_T \vert  x,B) + \partial_{B}   \ln P_{T,t}( x_T,B_T \vert  x,B) \right] 
\nonumber \\
&& =  \frac{\sqrt{2D} }{ [1-c^2(T,t)] } 
\left[     \frac{c(T,t)}{v(T,t)} 
 - \frac{  \sqrt{2D} e^{-(T-t)}}{\sigma(T,t)}   
   \right]     \left( \frac{  x e^{-(T-t)} -x_T}{\sigma(T,t)}\right) 
\nonumber \\
&& +  \frac{\sqrt{2D} }{ [1-c^2(T,t)] } 
\left[ \frac{  \sqrt{2D} c(T,t)  e^{-(T-t)}}{\sigma(T,t)}
 -  \frac{1}{v(T,t)}      
   \right]  \left( \frac{  B -B_T }{v(T,t)} \right)
\label{forcesupforwardOUexpli}
\end{eqnarray}
while the time-dependence is governed by the functions introduced in Eqs \ref{var}
and \ref{correl}
\begin{eqnarray}
\sigma(T,t)      && = \sqrt{  D \left[ 1- e^{- 2(T-t)}\right] }
     \nonumber \\
v(T,t)       && =     \sqrt{ T-t }
 \nonumber \\
c(T,t) &&= \sqrt{  \frac{  2    \left[ 1- e^{- (T-t)}\right]}{  (T-t) \left[ 1+ e^{- (T-t)}\right] } }
\label{timedep}
\end{eqnarray}

The explicit time-dependent force $F^{Forw[ x_T,B_T;T]}_t[ x ,B  ]  $ of Eq. \ref{forcesupforwardOUexpli}
can be then plugged into the stochastic differential equation 
of Eq. \ref{additivestar1d} involving a Wiener process $W(t)$
\begin{eqnarray}
dB^*(t)   =     \frac{ \left[ x^* (t)  dt +   dx^*(t) \right] }{ \sqrt{2D}  } 
 = \frac{   F^{Forw[\vec x_T,B_T;T]}[ x^*(t),B^*(t) ; t ]    }{ \sqrt{2D}  } dt  +dW(t)
\label{additivestar1dOU}
\end{eqnarray}
while the stochastic differential equation for $x^*(t)$ can be written either as 
Eq. \ref{langevinstar1d} 
\begin{eqnarray}
dx^*(t) =  \left(-  x^* (t)  + F^{Forw[\vec x_T,B_T;T]}[ x^*(t),B^*(t) ; t ] \right) dt + \sqrt{2D}   \ dW(t)
\label{langevinstar1dOU}
\end{eqnarray}
or as Eq. \ref{langevinstar1dsimpler}
\begin{eqnarray}
dx^*(t) = -  x^* (t)   dt + \sqrt{2D}  dB^*(t)
\label{langevinstar1dsimplerOU}
\end{eqnarray}
in order to generate stochastic trajectories of the conditioned process $( x^*(t),B^*(t))$.

As a final remark, let us stress that other explicit examples of microcanonical conditioning 
for the Brownian motion or the Ornstein-Uhlenbeck process
on various time-additive observables
can be found in \cite{Mazzolo_Brownian,Mazzolo_OU,bruyne_additive}.


\section{ Conclusion  }

\label{sec_conclusion}

In this paper, the recent studies concerning the conditioning of one-dimensional diffusion processes or discrete-time random walks on global dynamical constraints over a finite time-window $T$  \cite{Mazzolo_Brownian,Mazzolo_OU,bruyne_additive}
have been generalized to analyze the 'microcanonical conditioning' of Markov processes on time-additive observables. We have described the application to various types of Markov processes, namely discrete-time Markov chains, continuous-time Markov jump processes and diffusion processes in arbitrary dimension. In each setting, 
we have considered the most general time-additive observable that can involve both the time spent in each configuration and the elementary increments of the Markov process. We have illustrated the various cases via simple explicit examples. In the two Appendices, we describe the link with the 'canonical conditioning' based on the generating function of the time-additive observable, that has been much studied recently in the field of non-equilibrium steady states
\cite{peliti,derrida-lecture,tailleur,sollich_review,lazarescu_companion,lazarescu_generic,jack_review,vivien_thesis,lecomte_chaotic,lecomte_thermo,lecomte_formalism,lecomte_glass,kristina1,kristina2,jack_ensemble,simon1,simon2,simon3,Gunter1,Gunter2,Gunter3,Gunter4,chetrite_canonical,chetrite_conditioned,chetrite_optimal,chetrite_HDR,touchette_circle,touchette_langevin,touchette_occ,touchette_occupation,derrida-conditioned,derrida-ring,bertin-conditioned,garrahan_lecture,Vivo,chemical,touchette-reflected,touchette-reflectedbis,c_lyapunov,previousquantum2.5doob,quantum2.5doob,quantum2.5dooblong,c_ruelle,lapolla,chabane}
as recalled in the Introduction.

We hope that the present general formulation of the 'microcanonical conditioning' of Markov processes on time-additive observables will be helpful to identity new soluble cases besides the various explicit solutions given in the recent works 
\cite{Mazzolo_Brownian,Mazzolo_OU,bruyne_additive}.


\appendix


\section{ Links with the canonical conditioning on a time-additive observable for finite time $T$ }

\label{app_genecano}

As recalled in more details in the Introduction, 
the 'canonical conditioning' of Markov processes has been much studied recently 
in the field of non-equilibrium steady states
\cite{peliti,derrida-lecture,tailleur,sollich_review,lazarescu_companion,lazarescu_generic,jack_review,vivien_thesis,lecomte_chaotic,lecomte_thermo,lecomte_formalism,lecomte_glass,kristina1,kristina2,jack_ensemble,simon1,simon2,simon3,Gunter1,Gunter2,Gunter3,Gunter4,chetrite_canonical,chetrite_conditioned,chetrite_optimal,chetrite_HDR,touchette_circle,touchette_langevin,touchette_occ,touchette_occupation,derrida-conditioned,derrida-ring,bertin-conditioned,garrahan_lecture,Vivo,chemical,touchette-reflected,touchette-reflectedbis,c_lyapunov,previousquantum2.5doob,quantum2.5doob,quantum2.5dooblong,c_ruelle,lapolla,chabane}.
In this Appendix, it is thus interesting to describe the links with the microcanonical conditioning considered in
the main text.

\subsection{ Generating function $Z_{t,t_0}^{[k]}(x \vert x_0) $ of the total increment $A(t)-A(t_0)={\cal A}[x(t_0 \leq s \leq t) ] $  }

Here the basic object is the generating function $Z_{t,t_0}^{[k]}(x \vert x_0) $ of the total increment $A(t)-A(t_0)={\cal A}[x(t_0 \leq s \leq t) ] $
over the Markov trajectories $x(t_0 \leq s \leq t) $ starting at $x(t_0)=x_0$ and ending at $x(t)=x$
\begin{eqnarray}
Z_{t,t_0}^{[k]}(x \vert x_0) \equiv \langle \delta_{x(t),x} \ e^{k {\cal A}[x(t_0 \leq s \leq t) ]}  \ \delta_{x(t_0),x_0} \rangle
\label{genedef}
\end{eqnarray}
For fixed $k$, the generating function $Z_{t,t_0}^{[k]}(x \vert x_0) $ satisfies 

(i) some forward $k$-dependent dynamics with respect to the final state $x$ at time $t$,
that can be obtained from the forward dynamics of the joint propagator $P_{t,t_0}(x,A \vert x_0,A_0) $ of Eq. \ref{jointpropa} via
\begin{eqnarray}
Z_{t,t_0}^{[k]}(x \vert x_0) = \sum_A  e^{k A } P_{t,t_0}(x,A \vert x_0,A_0=0) 
\label{genejointforward}
\end{eqnarray}

(ii) some backward $k$-dependent dynamics with respect to the initial state $x_0$ at time $t_0$,
that can be obtained from the backward dynamics of the joint propagator $P_{t,t_0}(x,A \vert x_0,A_0) $ of Eq. \ref{jointpropa} via
\begin{eqnarray}
Z_{t,t_0}^{[k]}(x \vert x_0)  = \sum_{A_0} e^{-k A_0 } P_{t,t_0}(x,A=0 \vert x_0,A_0)
\label{genejointbackward}
\end{eqnarray}

Here it is important to stress that these two dynamics are not probability-conserving Markov dynamics,
since $Z_{t,t_0}^{[k]}(x \vert x_0) $ is a generating function and not a probability.


\subsubsection{ Dynamics of the generating function $Z_{t,t_0}^{[k]}(x \vert x_0) $ for discrete-time Markov chains of section \ref{sec_chain}   }

The joint generator $W(x,A ; x', A') $ of Eq. \ref{markovmatrixA} is in correspondence with the $k$-tilted matrix
\begin{eqnarray}
W^{[k]}(x; x') = \sum_A W(x,A ; x', A') e^{k (A-A') } = W(x ; x') e^{k \beta(x,x') }
\label{markovmatrixAk}
\end{eqnarray}
  
(i) The forward dynamics of Eq. \ref{markovchainAforward}
for the joint propagator $P_{t,t_0}(x,A \vert x_0,A_0) $
translates into the following forward dynamics for the generating function $Z_{t,t_0}^{[k]}(x \vert x_0) $
via Eq. \ref{genejointforward}
\begin{eqnarray}
Z_{t+1,t_0}^{[k]}(x \vert x_0) =  \sum_{x'}  W^{[k]}(x ; x')  Z^{[k]}_{t,t_0}(x'\vert x_0 )
\label{markovchainAforwardgene}
\end{eqnarray}

(ii) The backward dynamics of Eq. \ref{markovchainAbackward}
for the joint propagator $P_{t,t_0}(x,A \vert x_0,A_0) $
translates into the following backward dynamics for the generating function 
via Eq. \ref{genejointbackward}
\begin{eqnarray}
Z_{t,t_0-1}^{[k]}(x \vert x_0)  =
  \sum_{x_0'}    Z_{t,t_0}(x\vert x_0' )W^{[k]}(x_0'; x_0)
\label{markovchainAbackwardgene}
\end{eqnarray}


\subsubsection{ Dynamics of the generating function $Z_{t,t_0}^{[k]}(x \vert x_0) $ for continuous-time Markov jump processes of section \ref{sec_jump}   }

The jump-drift dynamics of Eqs \ref{additivejump} and \ref{wjoint}
is in correspondence with the $k$-tilted matrix
\begin{eqnarray}
w^{[k]}(x; x) && \equiv k \alpha(x) - \sum_{x' \ne x } w(x';x)
\nonumber \\
w^{[k]}(x; x') && \equiv w(x ; x') e^{k \beta(x,x') } \ \ \ \ \ \ \ \ \ {\rm for } \ \ x \ne x'
\label{markovmatrixjumpAk}
\end{eqnarray}

(i) The forward dynamics of Eq. \ref{markovjumpAforward}
for the joint propagator $P_{t,t_0}(x,A \vert x_0,A_0) $
translates into the following forward dynamics for the generating function $Z_{t,t_0}^{[k]}(x \vert x_0) $
via Eq. \ref{genejointforward}
\begin{eqnarray}
\partial _t Z_{t,t_0}^{[k]}(x \vert x_0)  
 =   \sum_{x'  }    w^{[k]}(x; x')  Z_{t,t_0}^{[k]}(x' \vert x_0) 
\label{markovjumpAforwardgene}
\end{eqnarray}

(ii) The backward dynamics of Eq. \ref{markovjumpAbackward}
for the joint propagator $P_{t,t_0}(x,A \vert x_0,A_0) $
translates into the following backward dynamics for the generating function 
via Eq. \ref{genejointbackward}
\begin{eqnarray}
- \partial _{t_0}Z_{t,t_0}^{[k]}(x \vert x_0)
 =    \sum_{x_0' } Z_{t,t_0}^{[k]}(x \vert x_0') w^{[k]}(x_0';x_0)  
\label{markovjumpAbackwardgene}
\end{eqnarray}


\subsubsection{ Dynamics of the generating function $Z_{t,t_0}^{[k]}(\vec x \vert \vec x_0) $ for diffusion processes of section \ref{sec_diff}   }

(i)  The forward generator of Eq. \ref{fokkerplanckAgenerator}
corresponds to the $k$-tilted  differential operator 
\begin{eqnarray}
 {\cal F}_k
 =  k \alpha[\vec x]  
   -  \sum_{i=1}^d \left( \partial_i  -k  \beta_i [ \vec x]  \right)   F_i[ \vec x ]  
   +  \sum_{i=1}^d   \left( \partial_i  -k  \beta_i [ \vec x]  \right)   D_i[ \vec x ]
 \left( \partial_i  - k  \beta_i [ \vec x]  \right)  
 \label{fokkerplanckAgeneratork}
\end{eqnarray}
The forward dynamics of Eq. \ref{fokkerplanckAforward}
for the joint propagator $P_{t,t_0}(\vec x,A \vert \vec x_0,A_0) $
translates for the generating function via Eq. \ref{genejointforward}
into the forward dynamics 
\begin{eqnarray}
 \partial_t Z_{t,0}^{[k]}(\vec x \vert \vec x_0) && \equiv {\cal F}_k Z_{t,0}^{[k]}(\vec x \vert \vec x_0)
  \nonumber \\
&& =    k \alpha[\vec x] Z_{t,0}^{[k]}(\vec x \vert \vec x_0)
   -  \sum_{i=1}^d \left( \partial_i  - k  \beta_i [ \vec x]  \right)   \bigg[ F_i[ \vec x ]  Z_{t,0}^{[k]}(\vec x \vert \vec x_0) \bigg]
 \nonumber \\
&&   +  \sum_{i=1}^d   
\left( \partial_i  - k  \beta_i [ \vec x]  \right)
\left[  D_i[ \vec x ]
 \left( \partial_i  - k  \beta_i [ \vec x]  \right)
 \bigg( Z_{t,0}^{[k]}(\vec x \vert \vec x_0)   \bigg) \right]
\label{fokkerplanckAforwardZ}
\end{eqnarray}

(ii) The backward dynamics of Eq. \ref{fokkerplanckAbackward} for the joint propagator 
$P_{T,t}(\vec x_T,A_T \vert \vec x,A) $
can be translated  for the generating function 
via Eq. \ref{genejointbackward}
into the backward dynamics
\begin{eqnarray}
 - \partial_t Z_{T,t}^{[k]}(\vec x_T \vert \vec x)  
 && =  k \alpha[\vec x] Z_{T,t}^{[k]}(\vec x_T \vert \vec x) 
   +  \sum_{i=1}^d F_i[ \vec x ]   \left( \partial_i  + k \beta_i [ \vec x]  \right)   Z_{T,t}^{[k]}(\vec x_T \vert \vec x) 
    \nonumber \\
&&   +   \sum_{i=1}^d   
 \left( \partial_i  + k \beta_i [ \vec x]  \right)
\bigg[  D_i[ \vec x ]
\left( \partial_i  + k \beta_i [ \vec x]  \right)Z_{T,t}^{[k]}(\vec x_T \vert \vec x)   \bigg]
  \nonumber \\
 && \equiv {\cal F}_k^{\dagger} Z_{t,0}^{[k]}(\vec x \vert \vec x_0)
\label{fokkerplanckAbackwardgene}
\end{eqnarray}
involving the adjoint operator of Eq. \ref{fokkerplanckAgeneratork}
\begin{eqnarray}
 {\cal F}_k^{\dagger}
 =  k \alpha[\vec x] 
   +  \sum_{i=1}^d F_i[ \vec x ]   \left( \partial_i  + k \beta_i [ \vec x]  \right)  
   +   \sum_{i=1}^d   
 \left( \partial_i  + k \beta_i [ \vec x]  \right)D_i[ \vec x ]
\left( \partial_i  + k \beta_i [ \vec x]  \right) 
  \label{fokkerplanckAgeneratorkdagger}
\end{eqnarray}


\subsection{ Conditional probability ${\cal P}^{Cond[k]}_t(x) $ 
if starting at $x_0$ at time $t=0$ and ending at $x_T$ at time $t=T$ }

Even if it is not a conserved probability, the generating function $ Z_{T,0}^{[k]}(x_T \vert x_0)$
satisfies nevertheless some analog of the Chapman-Kolmogorov Eq. \ref{chapman}
as a consequence of the additivity property of Eq. \ref{additivetraj}
\begin{eqnarray}
{\cal A}[x(0 \leq s \leq T) ] = {\cal A}[x(0 \leq s \leq t) ] + {\cal A}[x(t \leq s \leq T) ] 
\label{additivetwoparts}
\end{eqnarray}
that can be plugged into the definition of Eq. \ref{genedef} to obtain
\begin{eqnarray}
Z_{T,0}^{[k]}(x_T \vert x_0) && =
 \langle \delta_{x(T),x_T} e^{k {\cal A}[x(t \leq s \leq T) ]}\left[ \sum_x \delta_{x(t),x} \right]  \ e^{k {\cal A}[x(0 \leq s \leq t) ]}  \ \delta_{x(0),x_0} \rangle
 \nonumber \\
 && =\sum_x Z_{T,t}^{[k]}(x_T \vert x) Z_{t,0}^{[k]}(x \vert x_0)
\label{genechapman}
\end{eqnarray}
For each $k$, one can thus introduce the conditional probability $ {\cal P}^{Cond[k]}_t(x) $ 
to see the value $x$ at the internal time $t \in ]0,T[$ 
\begin{eqnarray}
{\cal P}^{Cond[k]}_t(x) =  \frac{Z_{T,t}^{[k]}(x_T \vert x) Z_{t,0}^{[k]}(x \vert x_0)}{Z_{T,0}^{[k]}(x_T \vert x_0)}
\label{markovcondk}
\end{eqnarray}
It is normalized as a consequence of Eq. \ref{genechapman}
\begin{eqnarray}
\sum_x {\cal P}^{Cond[k]}_t(x) = 1
\label{markovcondnormak}
\end{eqnarray}
and it satisfies the fixed boundary conditions at time $t=0$ and at time $t=T$
\begin{eqnarray}
{\cal P}^{Cond[k]}_0(x) && =  \frac{Z_{T,0}^{[k]}(x_T \vert x) Z_{0,0}^{[k]}(x \vert x_0)}{Z_{T,0}^{[k]}(x_T \vert x_0)} = \delta_{x,x_0}
\nonumber \\
{\cal P}^{Cond[k]}_T(x) && =  \frac{Z_{T,T}^{[k]}(x_T \vert x) Z_{T,0}^{[k]}(x \vert x_0)}{Z_{T,0}^{[k]}(x_T \vert x_0)} = \delta_{x,x_T}
\label{markovcondboundaryk}
\end{eqnarray}


\subsection{ Markov dynamics for the conditional probability ${\cal P}^{Cond[k]}_t(x) $  }

The Markov dynamics of the conditional probability ${\cal P}^{Cond[k]}_t(x)  $
can be derived
 from the Markov dynamics satisfied by the two generating functions in the numerator of Eq. \ref{markovcondk},
 namely :

(i) the forward dynamics of the generating function $Z_{t,0}^{[k]}(x \vert x_0) $ with respect to its final variable $x$ at time $t$

(ii) the backward dynamics of the generating function $Z_{T,t}^{[k]}(x_T \vert x) $ with respect to its initial variable $x$ at time $t$


\subsubsection{ Forward dynamics of the conditional probability ${\cal P}^{Cond[k]}_t(x) $ for discrete-time Markov chains of section \ref{sec_chain}   }

For the case of discrete-time Markov chains of section \ref{sec_chain}, the conditional probability
of Eq. \ref{markovcondk} 
satisfies
the forward dynamics
\begin{eqnarray}
{\cal P}^{Cond[k]}_{t+1}(x) =  \sum_{x'}  W^{Forw[k ; x_T,T]}_{t+1/2}(x ; x') {\cal P}^{Cond[k]}_t(x') 
\label{markovconddynk}
\end{eqnarray}
where the effective probabilities
\begin{eqnarray}
W^{{Forw[k ; x_T,T]}}_{t+1/2}(x ; x') = Z^{[k]}_{T,t+1}(x_T \vert x) W^{[k]}(x ; x')  \frac{ 1}{Z^{[k]}_{T,t}(x_T \vert x')} 
\label{markovcondforwardk}
\end{eqnarray}
involve the conjugation of the $k$-tilted matrix $W^{[k]}(x ; x') $ of Eq. \ref{markovmatrixAk}
with the generating functions 
$Z^{[k]}_{T,t+1}(x_T \vert x) $ and $Z^{[k]}_{T,t}(x_T \vert x')$ up to the imposed final value $x_T$ at time $T$.
Eq. \ref{markovcondforwardk} is the analog of Eq. \ref{markovcondforward} concerning the microcanonical conditioning.


\subsubsection{ Forward dynamics of the conditional probability ${\cal P}^{Cond[k]}_t(x) $ for continuous-time Markov jump processes of section \ref{sec_jump}   }

For the case of continuous-time Markov jump processes of section \ref{sec_jump},
the conditional probability
of Eq. \ref{markovcondk} 
satisfies
the forward dynamics
\begin{eqnarray}
&& \partial_t  {\cal P}^{Cond[k]}_t(x) 
 =  
  \sum_{x' \ne x }  \left[ w_t^{Forw[k;x_T,T]} (x ; x')   {\cal P}^{Cond[k]}_t( x') 
  - w_t^{Forw[k;x_T,T]} (x' ; x)   {\cal P}^{Cond[k]}_t( x)\right] 
\label{markovcondjumpforwardk}
\end{eqnarray}
where the effective rates 
\begin{eqnarray}
w_t^{Forw[k;x_T,T]} (x ; x') =    Z^{[k]}_{T,t}(x_T \vert x)  w^{k]}(x;x')
   \frac{1}{Z^{[k]}_{T,t}(x_T \vert x')} \ \ {\rm for } \ \ x \ne x'
\label{wjumpforwardk}
\end{eqnarray}
 involve the conjugation of the $k$-tilted matrix of Eq. \ref{markovmatrixjumpAk}
 with the generating functions 
$Z^{[k]}_{T,t+1}(x_T \vert x) $ and $Z^{[k]}_{T,t}(x_T \vert x')$ up to the imposed final value $x_T$ at time $T$.
Eq. \ref{wjumpforwardk} is the
 analog of Eq. \ref{wjumpforward} concerning the microcanonical conditioning.


\subsubsection{ Forward dynamics of the conditional probability ${\cal P}^{Cond[k]}_t(\vec x) $ for diffusion processes of section \ref{sec_diff}   }

For the case of diffusion processes of section \ref{sec_diff},
the conditional probability
of Eq. \ref{markovcondk} 
satisfies
the forward dynamics
\begin{eqnarray}
 \partial_t  {\cal P}^{Cond[k]}_t(\vec x) 
 =
    -  \sum_{i=1}^d  \partial_{x_i} 
\bigg[  \big( F_i[ \vec x ]  + F^{Forw[k ;\vec x_T,T]}_i[ \vec x ; t ]\big) {\cal P}^{Cond[k]}_t(\vec x)  \bigg]
   +  \sum_{i=1}^d   
 \partial_{x_i}  
\left[ D_i [\vec x]    \partial_{x_i} 
   {\cal P}^{Cond[k]}_t(\vec x)      \right] 
\label{markovconddiffderiforwardk}
\end{eqnarray}
where the additional time-dependent force 
\begin{eqnarray}
  F^{Forw[k ;\vec x_T,T]}_i[ \vec x ; t ] \equiv  
 2 D_i [\vec x]  \left[  k  \beta_i [ \vec x]          + \partial_{x_i}    \ln Z^{[k]}_{T,t}(\vec x_T \vert \vec x) \right]
\label{forcesupforwardk}
\end{eqnarray}
is the analog of Eq. \ref{forcesupforward} concerning the microcanonical conditioning.

The forward Fokker-Planck dynamics of Eq. \ref{markovconddiffderiforwardk}
can be translated into 
the following Stratonovich stochastic differential equations for the 
$d$ components $x_i^*(t)$ for $i=1,..,d$ in terms of $d$ independent Brownian motions $B_i(t)$ 
\begin{eqnarray}
dx_i^*(t) =  \left(f_i[ \vec x^* (t) ] + F^{Forw[k ;\vec x_T,T]}_i[ \vec x ; t ] \right) dt + g_i [\vec x^* (t) ]  \ dB_i(t)
\label{langevinstark}
\end{eqnarray}
where the only differences with respect to the unconditioned case of Eq. \ref{langevin}
are the additional time-dependent forces $F^{Forw[k ;\vec x_T,T]}_i[ \vec x ; t ] $ given in Eq. \ref{forcesupforwardk}.


\section{ Reminder on the conditioning for large $T$ when there is a normalizable steady state $P_{st}(x)$}

\label{app_largedev}

In this Appendix, the Markov processes $x(t)$ is assumed to
converge towards some normalizable steady-state $P_{st}(x)$.
This steady state $P_{st}(x)$ can be interpreted as the positive eigenvector $ \langle x \vert r_0 \rangle =r_0(x)$
associated to the highest eigenvalue of the Markov generator
\begin{eqnarray}
  P_{st}(x) =\langle x \vert r_0 \rangle = r_0(x)
\label{right0}
\end{eqnarray}
while the corresponding positive left eigenvector is constant 
\begin{eqnarray}
  \langle l_0 \vert x \rangle = l_0(x) = 1
\label{left0}
\end{eqnarray}
When the time interval $(t-t_0)$ becomes large, the propagator $P_{t,t_0}(x,x_0)$ is dominated by this highest eigenvalue contribution
\begin{eqnarray}
  P_{t,t_0}(x,x_0) \opsimeq_{(t-t_0) \to + \infty} \langle x \vert r_0 \rangle \langle l_0 \vert x_0 \rangle = r_0(x)  l_0(x) =P_{st}(x)
\label{domin0}
\end{eqnarray}
and describes the convergence towards the steady state $ P_{st}(x)$
for any initial condition $x_0$.

\subsection{ Asymptotic analysis of the generating function $Z_{t,t_0}^{[k]}(x \vert x_0) $ for large time interval $(t-t_0)$ }

For $k=0$, the generating function of Eq. \ref{genedef} coincides with the propagaor $  P_{t,t_0}(x,x_0) $ discussed above
\begin{eqnarray}
Z_{t,t_0}^{[k=0]}(x \vert x_0) =  P_{t,t_0}(x,x_0) 
\label{genek0}
\end{eqnarray}
As a consequence for $k \ne 0$, at least in some region around $k=0$, one expects that 
for large time-interval $(t-t_0)$, the generating function will be similarly dominated 
by the contribution of the highest eigenvalue of the appropriate $k$-deformed generator
\begin{eqnarray}
Z_{t,t_0}^{[k]}(x \vert x_0)  \opsimeq_{(t-t_0) \to + \infty} e^{(t-t_0) G(k)} \langle x \vert r_k \rangle \langle l_k \vert x_0 \rangle
= e^{(t-t_0) G(k)} r_k(x) l_k(x_0)
\label{geneddomin0}
\end{eqnarray}
with its positive right eigenvector $ r_k(x) \geq 0 $
and its positive left eigenvector $l_k(x) \geq 0$ satisfying the normalization
\begin{eqnarray}
1 = \langle l_k \vert r_k \rangle = \sum_x  \langle l_k \vert x \rangle  \langle x \vert r_k \rangle = \sum_x r_k(x) l_k(x)
 \label{Wktiltnorma}
\end{eqnarray}
while $[(t-t_0) G(k) ]$ represents the generating function of the cumulants of the time-additive observable $A_{t,t_0}$,
i.e. $ G(k)$ corresponds to the scaled cumulants generating function 
 in the large deviations theory, as recalled in more details below in subsection \ref{subsec_legendreinverse}.


\subsection{ Asymptotic analysis of the conditional probability $ {\cal P}^{Cond[k]}_t(x)$ at some interior time  $0 \ll t \ll T$}

For large $T$, if one is interested at some interior time $t$ satisfying $0 \ll t \ll T$,
one can plug the asymptotic behavior of Eq. \ref{geneddomin0}
into the three generating functions of Eq. \ref{markovcondk} to obtain the asymptotic behavior
of the conditional probability
\begin{eqnarray}
{\cal P}^{Cond[k]}_t(x) && \opsimeq_{0 \ll t \ll T}  \frac{ e^{(T-t) G(k)} r_k(x_T) l_k(x) e^{t G(k)} r_k(x) l_k(x_0)}
{e^{T G(k)} r_k(x_T) l_k(x_0)}
= l_k(x)  r_k(x)
\label{markovcondkdomin}
\end{eqnarray}
Since it is independent of the interior time $t$ as long as $0 \ll t \ll T$,
 it is useful to introduce the notation
\begin{eqnarray}
\rho_k (x) \equiv  l_k(x)  r_k(x)
 \label{rhokconditioned}
\end{eqnarray}
for the stationary density of the conditional probability ${\cal P}^{Cond[k]}_t(x) $ in the interior time region $0 \ll t \ll T$.


\subsection{ Physical meaning of the canonical $k$-conditioning in terms of the large deviations properties of $A(t)$ }

\label{subsec_legendreinverse}

Since the time-additive observable $A(t)$ of Eq. \ref{additivetraj} is extensive with respect to the time-interval,
it is useful to introduce its rescaled intensive counterpart
\begin{eqnarray}
a_{t,t_0} \equiv \frac{ A(t)-A(t_0) }{t-t_0} = \frac{ A[x(t_0 \leq s \leq t) ] }{t-t_0}
\label{additiveIntensive}
\end{eqnarray}
that will converge towards its steady value $a_{st}$ that can be computed from the steady state $P_{st}(x)$
and from the corresponding steady flows
\begin{eqnarray}
a_{t,t_0} \opsimeq_{(t-t_0) \to \infty} a_{st}
\label{g1}
\end{eqnarray}
The probability $P_{t,t_0}( a )  $ to see the value $a$ different from this steady value $a_{st} $
displays the large deviations form with respect to the time interval $(t-t_0)$
 \begin{eqnarray}
 P_{t,t_0}( a ) \opsimeq_{(t-t_0) \to +\infty} e^{- (t-t_0) I ( a )}
\label{level1def}
\end{eqnarray} 
where the positive rate function $I(a) \geq 0 $ vanishes only for the steady value $a_{st}$ of Eq. \ref{g1}
 \begin{eqnarray}
 I ( a_{st} ) =0
\label{iaeqvanish}
\end{eqnarray}
The generating function of the additive observable $A[x(t_0 \leq s \leq t) = A(t)-A(t_0)=(t-t_0) a_{t,t_0}$
can be evaluated from Eq. \ref{level1def}
via the saddle-point method for large $(t-t_0)$
\begin{eqnarray}
\langle e^{k A[x(t_0 \leq s \leq t) } \rangle = \langle e^{k (t-t_0) a_{t,t_0} } \rangle \equiv \int da  e^{k (t-t_0) a} P_{t,t_0}( a ) \opsimeq_{(t-t_0) \to +\infty} 
\int da e^{ (t-t_0) \left[ k a - I ( a ) \right] }\opsimeq_{(t-t_0) \to +\infty} e^{ (t-t_0) G(k) }
\label{level1gen}
\end{eqnarray} 
So the scaled cumulants generating function $G(k)$ that has been introduced in Eq. \ref{geneddomin0}
 is the Legendre transform of the rate function $ I ( a ) $
 \begin{eqnarray}
 k a - I ( a ) && =  G(k) 
 \nonumber \\
 k - I'(a) && =0
\label{legendre}
\end{eqnarray} 
while the reciprocal Legendre transform reads
 \begin{eqnarray}
 k a - G(k) && = I(a) 
 \nonumber \\
 a - G'(k) && =0
\label{legendrereci}
\end{eqnarray} 
As a consequence, the canonical $k$-conditioning discussed around Eq. \ref{rhokconditioned}
can be considered as asymptotically equivalent to the microcanonical conditioning on the intensive additive variable at the corresponding Legendre value $a=G'(k)$ of Eq. \ref{legendrereci}.


\subsection{ Corresponding time-independent generators of the conditioned dynamics for $1 \ll t \ll T$}

\label{subsec_doobleft}

\subsubsection{ Forward dynamics of the conditional probability ${\cal P}^{Cond[k]}_t(x) $ for discrete-time Markov chains of section \ref{sec_chain}   }

For the case of discrete-time Markov chains of section \ref{sec_chain}, 
the asymptotic form of Eq. \ref{geneddomin0} for the generating function
yields that the effective probabilities of Eq. \ref{markovcondforwardk} become
time-independent in the regime $1 \ll t \ll T $
\begin{eqnarray}
W^{{Forw[k ; x_T,T]}}_{t+1/2}(x ; x') &&  \opsimeq_{1 \ll t \ll T} 
e^{(T-t-1)  G(k)} r_k(x_T) l_k(x) W^{[k]}(x ; x')  \frac{ 1}{e^{(T-t) G(k)} r_k(x_T) l_k(x')} 
\nonumber \\
&&  \opsimeq_{1 \ll t \ll T} 
e^{ -  G(k)}  l_k(x) W^{[k]}(x ; x')  \frac{ 1}{  l_k(x')}
\label{markovcondforwardklargedev}
\end{eqnarray}
where $e^{G(k)}$ is the highest eigenvalue of the $k$-tilted matrix 
$W^{[k]}(x ; x')  $ of Eq. \ref{markovmatrixAk}, 
while $l_k(.) $ is the corresponding positive eigenvector 
\begin{eqnarray}
e^{G(k)} l_k(x') = \sum_x l_k(x) W^{[k]}(x ; x')
\label{eigenleftchain}
\end{eqnarray}
The corresponding positive right eigenvector $r_k(.)$
\begin{eqnarray}
e^{G(k)} r_k(x) = \sum_{x'} W^{[k]}(x ; x') r_k(x')
\label{eigenrightchain}
\end{eqnarray}
appears in the conditioned steady state of Eq. \ref{rhokconditioned} together with the left eigenvector  $l_k(.) $.


\subsubsection{ Forward dynamics of the conditional probability ${\cal P}^{Cond[k]}_t(x) $ for continuous-time Markov jump processes of section \ref{sec_jump}   }

For the case of continuous-time Markov jump processes of section \ref{sec_jump},
the asymptotic form of Eq. \ref{geneddomin0} for the generating function
yields that
 the effective rates of Eq. \ref{wjumpforwardk}  
 become
time-independent in the regime $1 \ll t \ll T $
\begin{eqnarray}
w_t^{Forw[k;x_T,T]} (x ; x') \opsimeq_{1 \ll t \ll T}     l_k(x)  w^{k]}(x;x')   \frac{1}{ l_k(x')}   \ \ {\rm for } \ \ x \ne x' 
\label{wjumpforwardklargedev}
\end{eqnarray}
where $l_k(.) $ is the positive eigenvector associated to the highest eigenvalue $G(k)$ of the $k$-tilted matrix 
$w^{[k]}(x ; x')  $ of Eq. \ref{markovmatrixjumpAk}
\begin{eqnarray}
G(k) l_k(x') = \sum_x l_k(x) w^{[k]}(x ; x') = l_k(x') w^{[k]}(x' ; x') +  \sum_{ x \ne x' } l_k(x) w^{[k]}(x ; x') 
\label{eigenleftjump}
\end{eqnarray}
Via the conservation of probability, the diagonal element can be computed in terms of the off-diagonal elements of Eq. \ref{wjumpforwardklargedev}
 using the eigenvalue Eq. \ref{eigenleftjump}
\begin{eqnarray}
w_t^{Forw[k;x_T,T]} (x' ; x') && = - \sum_{x \ne x'} w_t^{Forw[k;x_T,T]} (x ; x')
 \opsimeq_{1 \ll t \ll T} -  \left[ \sum_{x \ne x'}  l_k(x)  w^{k]}(x;x') \right]   \frac{1}{ l_k(x')} 
 \nonumber \\ &&
 = -  \left[ G(k) l_k(x') -  l_k(x') w^{[k]}(x' ; x')  \right]   \frac{1}{ l_k(x')} 
= w^{[k]}(x' ; x') - G(k)
\label{wjumpforwardklargedevdiag}
\end{eqnarray}
so that it involves the diagonal element $w^{[k]}(x' ; x') $ and the eigenvalue $G(k)$.

The positive right eigenvector $r_k(.)$ of the $k$-tilted matrix $w^{[k]}(x ; x')  $
\begin{eqnarray}
G(k) r_k(x) = \sum_{x'} w^{[k]}(x ; x') r_k(x')
\label{eigenrightjump}
\end{eqnarray}
appears in the conditioned steady state of Eq. \ref{rhokconditioned} together with the left eigenvector  $l_k(.) $.
 

\subsubsection{ Forward dynamics of the conditional probability ${\cal P}^{Cond[k]}_t(\vec x) $ for diffusion processes of section \ref{sec_diff}   }

For the case of diffusion processes of section \ref{sec_diff},
the asymptotic form of Eq. \ref{geneddomin0} for the generating function
yields that
the effective additional force of Eq. \ref{forcesupforwardk}
 becomes
time-independent in the regime $1 \ll t \ll T $
\begin{eqnarray}
  F^{Forw[k ;\vec x_T,T]}_i[ \vec x ; t ] \opsimeq_{1 \ll t \ll T}  
 2 D_i [\vec x]  \bigg(  k  \beta_i [ \vec x]          + \partial_{x_i}    \ln \big[ 
 e^{(T-t) G(k)} r_k( \vec x_T) l_k( \vec x)
 \big] \bigg)
 = 2 D_i [\vec x]  \bigg(  k  \beta_i [ \vec x]          + \partial_{x_i}    \ln \big[  l_k( \vec x)  \big] \bigg)
\label{forcesupforwardklargedev}
\end{eqnarray}
where $l_k( .) $ is the positive eigenvector associated to the highest eigenvalue $G(k)$ of the adjoint differential operator
$ {\cal F}_k^{\dagger} $ of Eq. \ref{fokkerplanckAgeneratorkdagger}
\begin{eqnarray}
G(k)   l_k( \vec x) && = {\cal F}_k^{\dagger}  l_k( \vec x)
\nonumber \\
 && =  k \alpha[\vec x] l_k( \vec x)
   +  \sum_{i=1}^d F_i[ \vec x ]   \left( \partial_i  + k \beta_i [ \vec x]  \right)  l_k( \vec x)
   +   \sum_{i=1}^d   
 \left( \partial_i  + k \beta_i [ \vec x]  \right) \left[ D_i[ \vec x ]
\left( \partial_i  + k \beta_i [ \vec x]  \right) l_k( \vec x) \right]
  \label{fokkerplanckAgeneratorkdaggereigen}
\end{eqnarray}
The corresponding positive eigenvector $r_k(.)$ of the operator ${\cal F}_k $ of Eq. \ref{fokkerplanckAgeneratorkdagger}
\begin{eqnarray}
G(k)   r_k( \vec x) && = {\cal F}_k  r_k( \vec x)
\nonumber \\
 &&
 =  k \alpha[\vec x]  r_k( \vec x)
   -  \sum_{i=1}^d \left( \partial_i  -k  \beta_i [ \vec x]  \right)  \left[ F_i[ \vec x ]  r_k( \vec x) \right]
   +  \sum_{i=1}^d   \left( \partial_i  -k  \beta_i [ \vec x]  \right) \left[  D_i[ \vec x ]
 \left( \partial_i  - k  \beta_i [ \vec x]  \right)  r_k( \vec x) \right]
\label{eigenrightdiff}
\end{eqnarray}
appears in the conditioned steady state of Eq. \ref{rhokconditioned} together with the left eigenvector  $l_k(.) $.


\end{document}